\documentclass[a4paper,12pt]{article}
\usepackage{amsfonts}
\usepackage{latexsym}
\usepackage{amssymb}
\date{}

\begin{document}

\title{Entropy Identity inducing\\
Non-Equilibrium Thermodynamics of\\Relativistic Multi-Component
Systems\\and their Newtonian Limits\thanks{In memory of Robert Trostel}
}
\author{W. Muschik\footnote{Corresponding author:
muschik@physik.tu-berlin.de}
\\
Institut f\"ur Theoretische Physik\\
Technische Universit\"at Berlin\\
Hardenbergstr. 36\\D-10623 BERLIN,  Germany}
\maketitle

            \newcommand{\be}{\begin{equation}}
            \newcommand{\beg}[1]{\begin{equation}\label{#1}}
            \newcommand{\ee}{\end{equation}\normalsize}
            \newcommand{\bee}[1]{\begin{equation}\label{#1}}
            \newcommand{\bey}{\begin{eqnarray}}
            \newcommand{\byy}[1]{\begin{eqnarray}\label{#1}}
            \newcommand{\eey}{\end{eqnarray}\normalsize}
            \newcommand{\beo}{\begin{eqnarray}\normalsize}
         
            \newcommand{\R}[1]{(\ref{#1})}
            \newcommand{\C}[1]{\cite{#1}}

            \newcommand{\mvec}[1]{\mbox{\boldmath{$#1$}}}
            \newcommand{\x}{(\!\mvec{x}, t)}
            \newcommand{\m}{\mvec{m}}
            \newcommand{\F}{{\cal F}}
            \newcommand{\n}{\mvec{n}}
            \newcommand{\argm}{(\m ,\mvec{x}, t)}
            \newcommand{\argn}{(\n ,\mvec{x}, t)}
            \newcommand{\T}[1]{\widetilde{#1}}
            \newcommand{\U}[1]{\underline{#1}}
            \newcommand{\V}[1]{\overline{#1}}
            \newcommand{\ub}[1]{\underbrace{#1}}
            \newcommand{\X}{\!\mvec{X} (\cdot)}
            \newcommand{\cd}{(\cdot)}
            \newcommand{\Q}{\mbox{\bf Q}}
            \newcommand{\p}{\partial_t}
            \newcommand{\z}{\!\mvec{z}}
            \newcommand{\bu}{\!\mvec{u}}
            \newcommand{\rr}{\!\mvec{r}}
            \newcommand{\w}{\!\mvec{w}}
            \newcommand{\g}{\!\mvec{g}}
            \newcommand{\D}{I\!\!D}
            \newcommand{\se}[1]{_{\mvec{;}#1}}
            \newcommand{\sek}[1]{_{\mvec{;}#1]}}            
            \newcommand{\seb}[1]{_{\mvec{;}#1)}}            
            \newcommand{\ko}[1]{_{\mvec{,}#1}}
            \newcommand{\ab}[1]{_{\mvec{|}#1}}
            \newcommand{\abb}[1]{_{\mvec{||}#1}}
            \newcommand{\td}{{^{\bullet}}}
            \newcommand{\eq}{{_{eq}}}
            \newcommand{\eqo}{{^{eq}}}
            \newcommand{\f}{\varphi}
            \newcommand{\rh}{\varrho}
            \newcommand{\dm}{\diamond\!}
            \newcommand{\seq}{\stackrel{_\bullet}{=}}
            \newcommand{\st}[2]{\stackrel{_#1}{#2}}
            \newcommand{\om}{\Omega}
            \newcommand{\emp}{\emptyset}
            \newcommand{\bt}{\bowtie}
            \newcommand{\btu}{\boxdot}
            \newcommand{\tup}{_\triangle}
            \newcommand{\tdo}{_\triangledown}
            \newcommand{\Ka}{\frac{\nu^A}{\Theta^A}}
            \newcommand{\K}[1]{\frac{1}{\Theta^{#1}}}
            \newcommand{\ap}{\approx}
            \newcommand{\bg}{\st{\Box}{=}}
            \newcommand{\si}{\simeq}
\newcommand{\Section}[1]{\section{\mbox{}\hspace{-.6cm}.\hspace{.4cm}#1}}
\newcommand{\Subsection}[1]{\subsection{\mbox{}\hspace{-.6cm}.\hspace{.4cm}
\em #1}}

\newcommand{\const}{\textit{const.}}
\newcommand{\vect}[1]{\underline{\ensuremath{#1}}}  
\newcommand{\abl}[2]{\ensuremath{\frac{\partial #1}{\partial #2}}}

\noindent
{\bf Keywords} Special-relativistic multi-component systems $\cdot$ Entropy Identity $\cdot$ 
Entropy balance of a component of the mixture $\cdot$ Entropy balance of the
mixture $\cdot$ Multi-heat relaxation $\cdot$ Equilibrium conditions: 4-temperature's Killing
relation $\cdot$ Newtonian limits of the balances of multi-component systems and their mixture

\vspace{1cm}\noindent
{\bf Abstract} Non-equilibrium and equilibrium thermodynamics of an interacting component in a special-relativistic multi-component system is discussed by use of an
entropy identity. The special case of the corresponding free component is considered.
Equilibrium conditions and especially the multi-component Killing relation of the
4-temperature are discussed. Two axioms characterize the mixture: additivity of the
energy momentum tensors and of the 4-entropies of the components generating those
of the mixture. The resulting quantities of a component and of the mixture, energy,
energy flux, momentum flux, stress tensor, entropy, entropy flux, supply and production
and their Newtonian limits in zeroth approximation are derived.

\section{Introduction}

The treatment of multi-component systems is often restricted to transport phenomena in
chemically reacting systems, that means, the mixture consisting of different components is
shortly
described by 1-component quantities such as temperature, pressure and energy which are not retraced to the corresponding quantities of the several components of the
multi-component system. That is the case in
non-relativistic physics \C{dGM} as well as in relativistic physics \C{K1,K2,IS,HASW,NEU}. In this paper, the single component as an interacting member of the mixture is investigated. Thus, each component of the mixture is equipped with its own temperature,
pressure, energy and mass density which all together generate the corresponding
quantities of the mixture.
\vspace{.3cm}\newline
Considering a multi-component system,  three items have to be distinguished: one component
as a member of the multi-component system which interacts with all the other components
of the system, the
same component as a free 1-component system separated from the multi-component
system and finally the multi-component system itself as a mixture which is composed of its
components. Here, all three items are discussed in a special-relativistic framework. For
finding out the entropy-flux, -supply, -production and -density, a special tool is used: the
entropy identity which constrains the possibility of an arbitrary choice of these quantities
\C{BOCH,MUBO,MUBO1}. Following J. Meixner and J.U. Keller that entropy in
non-equilibrium cannot be defined unequivocally \C{MEI,MEI1,KEL,KER}, the
entropy identity is only an (well set up) ansatz  for constructing a non-equilibrium
entropy and further corresponding quantities. This fact in mind,
a specific entropy and the corresponding Gibbs and
Gibbs-Duhem equations are derived. The definition of the energy-momentum tensor
and of the rest mass flux density are taken into account as contraints in the entropy
identity by introducing fields of Lagrange factors. The physical dimensions of these factors
allow to determine their physical meaning. The diffusion flux density is also introduced 
into the exploitation scheme.
\vspace{.3cm}\newline
Equilibrium is defined by equilibrium conditions which are divided into basic ones given
by vanishing entropy-flux, -supply and -production and into supplementary ones such as vanishing diffusion flux, vanishing heat flux and rest mass production. The Killing relation
of the 4-temperature concerning equilibrium is shortly discussed. Constitutive equations
are out of scope of this paper. 
\vspace{.3cm}\newline
The paper is organized as follows: After this introduction, the kinematics of a
multi-component system is considered in the next two sections for introducing the mass
flux and the diffusion flux densities. The energy-momentum tensor is decomposed
into its (3+1)-split, and the entanglement of the energy and momentum balances
are discussed, follwed by non-equilibrium thermodynamics of an interacting component
of the mixture and that of the corresponding free component. The equilibrium of
both is considered. Thermodynamics of the mixture starts with
two axioms: additivity of the energy momentum tensors and of the 4-entropies of the
components resulting in those of the mixture. Entropy, entropy flux, -supply and
-production are found out. The paper finishes with the Newtonian limit of  of the considered
relativistic thermodynamics in zeroth approximation, with a summary and an appendix..

\section{Kinematics}
\subsection{The components}

We consider a multi-component system consisting of $Z$ components.
The component index $^A$ runs from $1$ to $Z$. Each component has
its own rest frame ${\cal B}^A$ in which the rest mass density $\varrho^A$
is locally defined. These rest mass densities are relativistic invariants and therefore frame
independent\footnote{more details in Appendix \ref{RMD}}.
\vspace{.3cm}\newline
In general, the components have different 4-velocities: $u^A_k,\ A=1,2,...,Z;
k=1,...,4,$
which all are tensors of first order under Lorentz transformation. We now define the
component mass flux density as a 4-tensor of first order and the component mass production term as a scalar 
\bee{K2}
N^A_k\ :=\ \varrho^A u^A_k,\qquad N^{Ak}{_{,k}}\ =\ \Gamma^A.
\ee
Here, \R{K2}$_2$ is the mass balance equation of the $^A$-component .
Consequently, we introduce the basic fields of the components
\bee{K3}
\{\varrho^A,\ u^A_k\},\qquad A=1,2,...,Z.
\vspace{.3cm}\ee
The mass production term has two reasons: an external one by mass supply and one internal one
by chemical reactions
\bee{K3a} 
\Gamma^A\ =\ ^{(ex)}\Gamma^A + ^{(in)}\!\Gamma^A.
\ee
The external mass supply $^{(ex)}\Gamma^A$ depends on the environment of the system,
whereas $^{(in)}\Gamma^A$ is determined by chemical reactions depending on the
set of frame-independent stoichiometric equations which are discussed in Appendix \ref{SE}.

\subsection{The mixture\label{MIX}}

As each component, also the multi-component system has a mass density
$\varrho$ and a 4-velocity $u_k$ which are determined by the partial
quantities of the components. For deriving $\varrho$ and $u_k$, we apply
the nearly self-evident
\vspace{.3cm}\newline
$\blacksquare$\ {\sf Mixture Axiom:} The balance equation of a mixture
looks like the balance equation of an one-component system.
\hfill$\blacksquare$
\vspace{.3cm}\newline
Especially here, the mixture axiom is postulated for the balance equations of
mass, energy-momentum and entropy.
According to the mixture axiom, the mass balance of the mixture looks according to
\R{K2}$_2$  
\bee{K4}
N^k{_{,k}}\ =\ \Gamma,\qquad\Gamma\ =\ 0,
\ee
with vanishing total mass production, if the mass of the mixture is conserved\footnote{the
mixture as a closed system}. 
\vspace{.3cm}\newline
Now the question arises: which
quantities of the components of the mixture are additive? Obviously, neither the mass
densities $\varrho^A$ nor the 4-velocities $u^A_k$ are additive
quantities according to their definitions.
Consequently, we demand in accordance with the mixture axiom
that the mass flux densities are additive\footnote{The sign $\st{\td}{=}$ stands
for a setting and $:=$ for a definition.}
\bey\nonumber
\mbox{\sf Setting I:}\hspace{9.9cm}
\\ \label{K5}
\sum_A N^A_k\ \st{\td}{=}\ N_k\ :=\  \varrho u_k\ =\
\sum_A \varrho^A u^A_k\quad\longrightarrow\quad
u_k\ =\ \sum_A\frac{\varrho^A}{\varrho}u^A_k.
\eey
For the present, $\varrho$ and $u_k$ are unknown. Of course, they depend
on the basic fields of the components \R{K3}.
Contraction with $u^k$ and use of \R{K5}$_{2,3}$ results in
\bee{K6}
\varrho\ =\ \frac{1}{c^2}\sum_A\varrho^A u^A_ku^k\ =\ \frac{1}{c^2}N_ku^k\
=\ \frac{1}{c^2}N_k\frac{1}{\varrho}N^k\ \longrightarrow\
\varrho\ =\ \pm\frac{1}{c}\sqrt{N_kN^k},
\ee
or in more detail
\bee{K6a}
\varrho\ =\ \pm\frac{1}{c}\sqrt{\sum_{A,B}\varrho ^A\varrho^Bu^A_ku^{Bk}}.
\ee
The mass density $\varrho$ and the 4-velocity $u_k$ of the mixture are
expressed by those of the components according to \R{K6a} and \R{K5}$_4$.
According to \R{K5}$_4$, the 4-velocity of the mixture is a weighted mean value
of the 4-velocities of the components. For the mass density, we have according to
\R{K6}$_1$ also a with the Kluitenberg factor $f^A$ weighted mean value of the mass
density components \C{KLGR}
\bee{K6b}
f^A\ := \frac{1}{c^2}u^A_ku^k\quad\longrightarrow\quad
\rh\ =\ \sum_Af^A\rh^A\ =\ \sum_Af^A(u^A_k,u^k)\varrho^A,
\ee
resulting in the entanglement of $\rh$ and $u_k$ which are not independent of each
other
\bee{K6c}
\rh\ =\ R(\rh^A,u^A_k,u_k),\qquad u_k\ =\ U_k(\rh^A,u^A_k,\rh).
\vspace{.3cm}\ee
According to \R{K5}$_1$ and \R{K2}$_2$, we obtain the additivity of the mass
production terms
\bee{K7a}
N^k{_{,k}}\ =\ \sum_AN^{Ak}{_{,k}}\ =\ \sum_A\Gamma ^A\ =\ \Gamma
\ \longrightarrow\ \sum_A {^{ex}}\Gamma ^A\ =\ ^{ex}\Gamma,\
\sum_A {^{in}}\Gamma ^A\ =\ 0.\footnote{Chemical reactions are mass conserving.}
\ee

\subsection{The diffusion flux}

From \R{K5}$_3$ and \R{K6b}$_2$ follows
\bee{M1}
0\ =\ \sum_A\rh^Au^A_k - u_k\sum_Af^A\rh^A\ =\ \sum_A \rh^A(u^A_k-f^Au_k).
\ee
Introducing the diffusion flux density using \R{M1}$_2$
\bee{K8c}
J^A_k\ :=\ \varrho^A(u^A_k - f^Au_k)\ =\ N^A_k - \rh^Af^Au_k
\quad\longrightarrow\quad
\sum_A J^A_k\ =:\ J_k\ =\ 0,
\ee
we obtain 
\byy{K9}
J^A_ku^k &=& \varrho^A(u^A_ku^k - f^Ac^2)\ =\ 0,
\\ \label{K9a}
J^A_ku^{Ak}&=& c^2\rh^A[1-(f^A)^2]\ =:\ c^2\rh^Aw^A\
=\ w^AN^A_ku^{Ak},
\\ \label{K9b}
1-w^A\ \geq\ 0.\hspace{-7.3cm}
\eey
By introducing the projectors
\bee{K15}
h^{Am}_l\ :=\ \delta^m_l - \frac{1}{c^2}u^{Am}u^{A}_l,\qquad
h^{m}_l\ :=\ \delta^m_l - \frac{1}{c^2}u^{m}u_l,
\ee
we obtain the following properties of the diffusion flux density:
\byy{L1}
J^{Am}h^k_m &=& J^{Ak}\ =\ N^{Am} h^k_m
\\ \label{L2}
J^{Am}h^{Ak}_m &=& \rh^Af^A(f^Au^{Ak}-u^k)
\\ \label{L3}
J^{Ak} &=& J^{Am}h^{Ak}_m + \rh^Aw^Au^{Ak}\ =\
J^{Am}h^{Ak}_m+w^AN^{Ak}                                                 
\\ \label{K8d1}
J^{Ak}{_{,k}} &=&(J^{Am}h^{Ak}_m)_{,k}
+(\rh^Aw^A){_{,k}}u^{Ak}+\rh^Aw^Au^{Ak}{_{,k}}.
\eey
According to \R{L1}$_2$, the diffusion flux density is that part of the mass flux density
which is perpendicular to the 4-velocity of the mixture. The diffusion flux density
vanishes in 1-component systems ($u^A_k\equiv u_k$) according to $f^A=f=1$ and \R{K8c}$_1$.

\section{The Energy-Momentum Tensor}
\subsection{Free and interacting components\label{FIC}}

The energy-momentum tensor $T^{Akl}$ of the $^A$-component consists of two parts
\bee{O1}
T^{Akl}\ =\ \st{0}{T}\!{^{Akl}} + \sum_B W^{Akl}_B,\quad W^{Bkl}_B\ =\ 0.
\ee
Here, $\st{0}{T}\!{^{Akl}}$ is the energy-momentum tensor of the free
$^A$-component, that is the case, if there are no interactions between the
$^A$-component and the other ones. $W^{Akl}_B$ describes the interaction
between the $^B$- and the $^A$-component.
The interaction between the external environment and the $^A$-component is given by the force density $k^{Al}$ which appears in the energy-momentum balance equation
\bee{O2}
T^{Akl}{_{,k}}\ =\ k^{Al}\ =\ \Omega^{Al}+\frac{1}{c^2}u^{Al}u^A_mk^{Am},
\qquad \Omega^{Al}u^A_l\ =\ 0,
\ee
and in the balance equations of
\byy{K13}
\mbox{energy:}\hspace{1.4cm}
u^A_lT^{Akl}{_{,k}} &=& u^A_lk^{Al}\ =:\ \Omega^A,
\\ \label{K14}
\mbox{and momentum:}\hspace{.5cm}
h^{Am}_lT^{Akl}{_{,k}} &=& h^{Am}_lk^{Al}\ =:\ \Omega^{Am}.
\eey
Consequently, the interaction of the $^A$-component with the other components
of the mixture modifies the energy-momentum tensor of the free $^A$-component.
Additionally, its interaction with the environment shows up in the source of
the energy-momentum balance. According to its definition, $T^{Akl}$ is the energy-momentum tensor of the "$^A$-component in the mixture".

\subsection{(3+1)-split}

The (3+1)-split of the energy-momentum tensor of the $^A$-component is
\bee{J1}
T^{Akl}\ =\ \frac{1}{c^2}e^Au^{Ak}u^{Al} + 
u^{Ak}p^{Al} +\frac{1}{c^2}q^{Ak}u^{Al} +t^{Akl}.
\ee
The (3+1)-components of the energy-momentum tensor are\footnote{the (3+1)-split
is made by taking the physical meaning of \R{J2} and \R{J3} into account, see \R{J6} to \R{J8a}}
\byy{J2}
e^A\ :=\ \frac{1}{c^2}T^{Ajm}u^A_ju^A_m ,\qquad
p^{Al}\ :=\ \frac{1}{c^2}h^{Al}_mT^{Ajm}u^A_j,
\\ \label{J3}
q^{Ak}\ :=\ h^{Ak}_jT^{Ajm}u^A_m,\qquad
t^{Akl}\ :=\ h^{Ak}_jT^{Ajm}h^{Al}_m,
\\ \label{J3a}
q^{Ak}u^A_k = 0,\ p^{Al}u^A_l = 0,\quad t^{Akl}u^A_k =0,\ t^{Akl}u^A_l = 0.
\vspace{.3cm}\eey
The symmetric part of the energy-momentum tensor \R{J1} is
\bee{J3b}
T^{A(kl)}\ =\ \frac{1}{c^2}e^Au^{Ak}u^{Al}
+\frac{1}{2c}u^{Ak}\Big(cp^{Al}+\frac{1}{c}q^{Al}\Big)
+\frac{1}{2c}\Big(cp^{Ak}+\frac{1}{c}q^{Ak}\Big)u^{Al}+t^{A(kl)},
\ee
and its anti-symmetric part is
\bee{J3c}
T^{A[kl]}\ =\ \frac{1}{2c}u^{Ak}\Big(cp^{Al}-\frac{1}{c}q^{Al}\Big)
-\frac{1}{2c}\Big(cp^{Ak}-\frac{1}{c}q^{Ak}\Big)u^{Al}+t^{A[kl]}.
\ee
The stress tensor is composed of the pressure $p^A$ and the viscous tensor
$\pi^{Akl}$
\bee{cT2}
t^{Akl}\ =\ -p^Ah^{Akl} + \pi^{Akl},\qquad t^{Ak}_k\ =\ -3p^A.
\vspace{.3cm}\ee
We now consider the physical dimensions of the introduced quantities\footnote{the bracket [$\boxtimes$] signifies the physical dimension of $\boxtimes$}.
According to \R{K15} and \R{K6b}$_1$, we have
\bee{J5}
[h^{Al}_m]\ =\ 1,\qquad [f^A]\ =\ 1.
\ee
By taking \R{cT2}, \R{J5}$_1$ and \R{J1} into account we obtain
\byy{J6}
 [t^{Akl}]\ =\ [p^{A}]\ =\ [\pi^{Akl}]\ =\ [e^A]\ =\ [q^{Ak}]\frac{s}{m}&=& [p^{Al}]\frac{m}{s},
\\ \nonumber
\mbox{pressure}\ =\ [p^{A}]\ =\ \frac{N}{m^2}\ =\ \frac{Nm}{m^3}
&=& \mbox{energy density}\ =\hspace{1.2cm} 
\\ \label{J7}
=\ \frac{kg\ m}{s^2}\frac{1}{m^2}\
=\ kg\frac{m}{s}\frac{1}{m^3}\frac{m}{s}&=&\mbox{momentum flux density},
\\ \label{J8}
[q^{Ak}]\ =\ [e^A]\frac{m}{s}\ =\ \frac{Nm}{m^3}\frac{m}{s}&=&
\mbox{energy flux density},
\\ \label{J8a}
[p^{Al}]\ =\ kg\frac{m}{s}\frac{1}{m^3}&=&\mbox{momentum density}.
\vspace{.3cm}\eey
The (3+1)-split \R{J1} of the energy-momentum tensor can be written in a more
compact form 
\byy{K11a}
T^{Akl}\ =\ \frac{1}{c^2}Q^{Ak}u^{Al} + \tau^{Akl},\hspace{2.5cm}
\\ \label{K11b}
Q^{Ak}\ :=\ e^Au^{Ak} + q^{Ak},\qquad
\tau^{Akl}\ :=\ u^{Ak}p^{Al} + t^{Akl}.
\eey
The energy-momentum tensor \R{K11a} is that of the $^A$-component in the
mixture, that means as dicussed in sect.\ref{FIC}, the  $^A$-component is not a free system and the
(3+1)-split-components $e^A, q^{Ak}, p^{Al}$ and $ t^{Akl}$ include the internal
interaction of the $^A$-component with all the other ones.

\subsection{Additivity}

We now consider the equivalent-system composed of the $Z$  components: that is the mixture
which consists of these $Z$ interacting components. Because this interaction
is already taken into account by the (3+1)-split-components, the
energy-momentum tensors of the components are additive without
additional interaction terms. Consequently, the energy-momentum tensor ${\sf T}^{kl}$
of the mixture is
\bey\nonumber
\mbox{\sf Setting II:}\hspace{9.5cm}
\\ \label{K11c}
{\sf T}^{kl}\ :=\ \frac{1}{c^2}Q^{k}u^{l} + \tau^{kl}  \st{\td}{=}\
\sum_A T^{Akl}\ =\
\sum_A\Big(\frac{1}{c^2}Q^{Ak}u^{Al} + \tau^{Akl}\Big).
\eey
Multiplication with $u_l$ results by use of \R{K6b}$_1$ and \R{K11b}$_2$ in
\bee{K11d}
Q^k\ =\ \sum_A\Big(Q^{Ak}f^A + \tau^{Akl}u_l\Big),
\ee
and by multiplication with $h_l^m$,  \R{K11c} results in
\byy{K15a}
\tau^{km} &=& \sum_A\Big(Q^{Ak}g^{Am} +
\tau^{Akl}h_l^m\Big),
\\ \label{K15a1}
g^{Am} &:=& \frac{1}{c^2}u^{Al}h_l^m\ =\ 
\frac{1}{c^2}(u^{Am}-f^Au^m)\ =\ \frac{1}{c^2\rh^A}J^{Am}.
\eey
For an 1-component system ($u^A_k\equiv u_k$), we obtain according to \R{K15a1}
$g^{Am}=g^m=0$ taking $f^A=f=0$ into account.

\subsection{(3+1)-components of the mixture\label{COMI}}

Starting with \R{K11b}, we obtain
\byy{K15a2}
Q^{Ak}u^A_k\ =\ c^2{\sf e}^A,&\quad& Q^{Ak}h_k^{Am}\ =\ {\sf q}^{Am},
\\ \label{K15a3}
\tau^{Akm}u^A_k\ =\ c^2{\sf p}^{Am},&\quad&
\tau^{Akm}h_k^{Aj}\ =\ {\sf t}^{Ajm}.
\eey
According to \R{K11a} and \R{K11c}$_1$, these relations are analogous for the mixture.
Consequently, from \R{K11d} follows
\bee{K15b}
Q^ku_k\ =:\ c^2{\sf e}\ =\ \sum_A\Big(Q^{Ak}f^Au_k +
\tau^{Akl}u_lu_k\Big),
\ee
resulting with \R{K15a2}$_1$ in the energy density of the mixture
\bee{K15c}
c^2{\sf e}\ =\ \sum_A\Big(c^2e^A(f^A)^2 + q^{Ak}f^Au_k +c^2p^{Al}f^Au_l +
t^{Akl}u_lu_k\Big).
\ee
From \R{K11d} follows
\bee{K15d}
Q^kh_k^m\ =:\ {\sf q}^m\ =\ \sum_A\Big(f^AQ^{Ak} h^m_k +
h^m_k\tau^{Akl}u_l\Big),
\ee
resulting with \R{K15a2}$_2$ in the energy flux density of the mixture
\bee{K15e}
{\sf q}^m\ =\ \sum_A\Big(c^2e^Af^Ag^{Am} + q^{Ak}f^Ah^m_k +
c^2p^{Al}g^{Am}u_l + t^{Akl}h^m_ku_l\Big).
\ee
From \R{K15a} follows
\bee{K15f}
\tau^{km}u_k\ =:\ c^2{\sf p}^m\ =\ \sum_A\Big(Q^{Ak}u_kg^{Am} +
\tau^{Akl}h^m_lu_k\Big),
\ee
resulting with \R{K15a3}$_1$ in the momentum density of the mixture
\bee{K15g}
c^2{\sf p}^m\ =\ \sum_A\Big(c^2e^Af^Ag^{Am} + q^{Ak}u_kg^{Am} +
c^2p^{Al}f^Ah^m_l + t^{Akl}h^m_lu_k\Big).
\ee
And from \R{K15a3}$_2$ follows finally
\bee{K15h}
\tau^{km}h_k^j\ =:\ {\sf t}^{jm}\ =\ \sum_A\Big(Q^{Ak}g^{Am}h^j_k +
\tau^{Akl}h^m_lh^j_k\Big)
\ee
which by taking \R{K11b} into account results in the stress tensor and the pressure of the
mixture
\byy{K15j}
{\sf t}^{jm} &=& \sum_A\Big(c^2e^Ag^{Aj}g^{Am} + q^{Ak}h^j_kg^{Am} +
c^2p^{Al}g^{Aj}h^m_l + t^{Akl}h^j_kh^m_l\Big),
\\ \nonumber
{\sf p} &=& -\frac{1}{3}{\sf t}^{jm}h_{jm}\ =\ 
-\frac{1}{3}\sum_A\Big(Q^{Ak}g^{Am}h^j_k + \tau^{Akl}h^m_lh^j_k\Big)h_{jm}\
=
\\  \nonumber
&\mbox{}&\hspace{2.1cm} =\ 
-\frac{1}{3}\sum_A\Big(\frac{1}{c^2}Q^{Ak}u^{Ap}h_{kp} + \tau^{Akl}h_{kl}\Big)\
=
\\ \label {K15j1}
&\mbox{}&\hspace{2.1cm}=\ 
-\frac{1}{3}\sum_A\Big(\frac{1}{c^2}( e^Au^{Ak} + q^{Ak})u^{Ap}h_{kp} +
(u^{Ak}p^{Al} + t^{Akl})h_{kl}\Big).
\vspace{.3cm}\eey
The additivity of the energy-momentum tensors \R{K11c}
results in \R{K15c}, \R{K15e}, \R{K15g} and \R{K15j}, relations which
express the (3+1)-components of the energy-momentum tensor of the
mixture as those of the components and their velocities
\byy{K15k}
\Big\{{\sf e},{\sf q}^k,{\sf p}^k,{\sf t}^{kl}\Big\} &=& F\Big(e^A,q^{Ak},p^{Ak},t^{Akl},
u^{Ak},\rh(\rh^A,u^A_k,u^k), u^k(\rh^A,u^A_k,\rh)\Big),
\\ \label{K15k1}
{\sf T}^{kl} &=& \frac{1}{c^2}{\sf e}u^{k}u^{l} + 
u^{k}{\sf p}^{l} +\frac{1}{c^2}{\sf q}^{k}u^{l} +{\sf t}^{kl},\qquad
{\sf t}^{kl}\ =\ -{\sf p}h^{kl}+^\dm\!\pi^{kl}.
\eey
The 4-velocity $u^k$ is given by \R{K5}$_4$.
\vspace{.3cm}\newline
The influence of the additivity of the energy-momentum tensors on the
balance equations of energy and momentum is investigated in the next section.

\section{Entanglement of Energy and Momentum Balances}

If the energy-momentum tensors of the $^A$-component and of the mixture
are $T^{Akl}$ and ${\sf T}^{kl}$, the energy and momentum balances are
according to the mixture axiom by use of \R{K13} and \R{K14}
\byy{K13z}
\mbox{energy:}\hspace{1.4cm}
u^A_lT^{Akl}{_{,k}}\ =\ \Omega^A
&\quad& u_l{\sf T}^{kl}{_{,k}}\ =\ \Omega,
\\ \label{K14z}
\mbox{momentum:}\hspace{.5cm}
h^{Am}_lT^{Akl}{_{,k}}\ =\ \Omega^{Am}
&\quad& h^m_l{\sf T}^{kl}{_{,k}}\ =\ \Omega^m.
\eey
The balances \R{K13z}$_3$ and \R{K14z}$_3$ follow from \R{K13} and \R{K14}
by the mixture axiom.
Here, $\Omega^A$ and  $\Omega$ are the energy supplies, and
$\Omega^{Am}$ and  $\Omega^m$ the momentum supplies of
the $^A$-component and of the mixture.
\vspace{.3cm}\newline
The (3+1)-split of the divergence of the energy-momentum tensor of the
$^A$-component results by use of \R{K15}$_1$ in
\bee{K14a}
\delta^m_lT^{Akl}{_{,k}}\ =\ T^{Akm}{_{,k}}\ =\ h^{Am}_lT^{Akl}{_{,k}}+
\frac{1}{c^2}u^{Am}u^A_lT^{Akl}{_{,k}}.
\ee
If the component index $^A$ is cancelled in \R{K14a}, we obtain
the decomposition of the divergence of the energy-momentum tensor of the mixture.
Taking \R{K13z} and \R{K14z} into account, these divergences can be written as
\bee{K14d}
T^{Akm}{_{,k}}\ =\ \Omega^{Am} + \frac{1}{c^2}u^{Am}\Omega^A,\qquad
{\sf T}^{km}{_{,k}}\ =\ \Omega^{m} + \frac{1}{c^2}u^m\Omega.
\vspace{.3cm}\ee
The additivity of the energy-momentum tensors \R{K11c} results in the additivity of
the force densities\footnote{this is a strong argument for the validity of Setting II
\R{K11c}}
\bee{K14e} 
k^m\ =\ \Omega^{m} + \frac{1}{c^2}u^m\Omega\ =\
\sum_A\Big(\Omega^{Am} + \frac{1}{c^2}u^{Am}\Omega^A\Big)\ =\
\sum_A k^{Am}.
\ee
Taking \R{K13}$_2$ and \R{K14}$_2$ into account, we obtain by
multiplication of \R{K14e} with $u_m$, resp. with $h^p_m$,
\bee{K14f}
\Omega\ =\ \sum_A\Big(\Omega^{Am}u_m + f^A\Omega^A\Big),
\qquad
\Omega^p\ =\ \sum_A\Big(\Omega^{Am}h^p_m + 
g^{Ap}\Omega^A\Big).
\ee
Inserting \R{K13} and \R{K14}, we obtain in more detail
\byy{K14g}
u_l{\sf T}^{kl}{_{,k}}\ =\ \sum_A\Big\{\Big(h^{Am}_lu_m +
f^Au^A_l\Big)T^{Akl}{_{,k}}\Big\},
\\ \label{K14h}
h^p_l{\sf T}^{kl}{_{,k}}\ =\ \sum_A\Big\{\Big(h^{Am}_lh^p_m +
g^{Ap}u^A_l\Big)T^{Akl}{_{,k}}\Big\}.
\eey
As \R{K14f} indicates, the additivity of the energy-momentum
tensors causes
that the supplies of energy and momentum are entangled, expressed
by the inequalities
\bee{K16}
\sum_A f^A\Omega^A\ \neq\ \Omega,\qquad
\sum_A \Omega^{Am}h^p_m\ \neq\ \Omega^p.
\vspace{.3cm}\ee
Also if the total force density and the total momentum supply are zero,
\bee{K17}
{\sf T}^{kl}{_{,k}}\ =\ 0\quad\longrightarrow\quad\Omega_{\sf iso}\ =\ 0\
\wedge\ \Omega^m_{\sf iso}\ =\ 0,
\ee
we obtain according to \R{K14f}$_{1,2}$
\byy{K18}
\sum_A\Omega^{Am}_{\sf iso}u_m &=&
-\sum_Af^A\Omega^A_{\sf iso}\ \neq\ 0,
\\ \label{K19}
\sum_A\Omega^{Am}_{\sf iso}h^p_m &=& 
-\sum_Ag^{Ap}\Omega^A_{\sf iso}\ \neq\ 0.
\eey
As expected,  the supplies of energy and momentum remain entangled in a system
of vani\-shing total force and momentum densities. The entanglement vanishes for such isolated
systems for which the force and momentum supplies for all $^A$-components are zero.

\section{Thermodynamics of Interacting Components\label{IC}}
\subsection{The entropy identity\label{IC1}}

For establishing the entropy balance equation, we use a special procedure
starting out with an identity \C{RELTAG,MUBO}, the so-called {\em entropy identity}.
This tool helps to restrict arbitrariness for defining entropy density, entropy flux density,
entropy production and supply. In the sequel, we establish an entropy identity 
for the $^A$-component by starting out with the (3+1)-split of the entropy 4-vector
\bee{T2}
S^{Ak}\ =\ s^Au^{Ak} + s^{Ak}.
\ee
Here $s^A$ is the entropy density and $s^{Ak}$ the entropy flux density defined by
\bee{T2z}
s^A\ :=\ \frac{1}{c^2}S^{Ak}u^A_k,\quad
s^{Ak}\ :=\ S^{Am}h^{Ak}_m.
\ee
From non-relativistic physics, we know the physical dimensions
\bee{T2w}
[s^A]\ =\ [e^A]\frac{1}{K}\ =\ \frac{Nm}{m^3}\frac{1}{K},\quad
[s^{Ak}]\ =\ [q^{Ak}]\frac{1}{K}\ =\
 \frac{Nm}{m^3}\frac{m}{s}\frac{1}{K}.
\vspace{.3cm}\ee
Before writing down the entropy identity, we have to choose the quantities which are
essential for formulating the four entropy quantities mentioned above. The choice is:
all quantities appearing in the (3+1)-split of the energy-momentum tensor \R{J1} and in the mass flux density \R{K2}$_1$ have to be included in the entropy identity which is generated by adding suitable zeros to \R{T2}. There is no unequivocal entropy identity
\C{KEL} and consequently, also no unique entropy density, -flux, -supply
and -production:
\bey\nonumber
S^{Ak}&\equiv& s^Au^{Ak} + s^{Ak}
+\kappa^A\Big(N^{Ak}-\rh^Au^{Ak}\Big)+
\\ \label{aT2}
&&+\ \Lambda ^A_l\Big(T^{Akl}-\frac{1}{c^2}e^Au^{Ak}u^{Al} - 
u^{Ak}p^{Al} -\frac{1}{c^2}q^{Ak}u^{Al} -t^{Akl}\Big).
\eey
The Lagrange factors $\kappa^A$ and $\Lambda^A_l$ are field functions whose physical
meaning becomes clear in the course of the exploitation of the entropy identity.
Here, $\kappa^A$ is a scalar, undefined for the present, and
the (3+1)-split of the likewise arbitrary vector $\Lambda^A_l$ is
\byy{bT2}
\Lambda^A_{l}\ =\ \lambda^Au^A_{l}+ \lambda^A_l,\qquad
\lambda^A_lu^{Al}\ =\ 0,&\quad& \lambda^A_j h^{Aj}_l\ =\ \lambda^A_l,
\\ \label{bT2a}
\Lambda^A_lu^{Al}\ =\ c^2\lambda^A,&\quad&\Lambda^A_lh^{Al}_m\ =\
\lambda^A_m. 
\eey
We denote the Lagrange factors
$\kappa^A$, $\lambda^A$ and $\lambda^A_l$ as {\em accessory variables} because
they help to formulate an entropy identity. An identification of these auxiliary variables
is given below after the definitions of entropy flux density, entropy production, density
and supply in section \ref{ACVA}.
By use of \R{bT2}, the entropy identity \R{aT2} becomes
\bey\nonumber
S^{Ak}\ \equiv\hspace{-.3cm}
&& u^{Ak}\Big(s^A
-\kappa^A\rh^A
-\lambda^Ae^A
-\lambda^A_lp^{Al}\Big)+\kappa^AN^{Ak}+
\hspace{2.3cm} 
\\ \label{dT2}
&&+\ s^{Ak}   
-\lambda^Aq^{Ak} 
-\lambda^A_lt^{Akl}
+\ \Big(\lambda^Au^A_{l}+ \lambda ^A_l\Big)T^{Akl}.
\eey
This identity transforms in an other one by differentiation and by taking the entropy
balance equation
\bee{T5}
S^{Ak}{_{,k}}\ =\ \sigma^A + \varphi^A 
\ee
into account.                                                                                     
\bey\nonumber
S^{Ak}{_{,k}}\ \equiv\hspace{-.3cm} 
&&\Big[u^{Ak}\Big(s^A 
-\kappa^A\rh^A
-\lambda^A e^A - \lambda^A_lp^{Al}\Big)\Big]_{,k}
+\ \Big[\ s^{Ak}
-\lambda^Aq^{Ak}
- \lambda^A_lt^{Akl}\Big]_{,k}+
\\ \nonumber
&&+\kappa^A\Gamma^A+\kappa^A{_{,k}}N^{Ak}+
\\ \label{T4}
&&+\Big(\lambda^A u^A_l\Big)_{,k}T^{Akl} + \lambda^A u^A_lT^{Akl}{_{,k}}+
\lambda^A{_{l,k}}T^{Akl}+ \lambda^A_{l}T^{Akl}{_{,k}}\ =\
\sigma^A + \varphi^A .
\eey
Here, $\sigma^A$ is the entropy production  and $\varphi^A$ the
entropy supply of the $^A$-component. The identity
\R{T4} changes into the entropy production, if $s^A$, $s^{Ak}$
and $\varphi^A$ are specified below.                           
\vspace{.3cm}\newline
Now we look for terms of the third row of \R{T4} which fit into the first row
of \R{T4}. The shape of these terms is $[u^{Ak}{\sf scalar}]_{,k}$ according to the
first term of \R{T4} and $\Psi^{Ak}{_{,k}}\
(\Psi^{Ak}u^A_k=0) $ according to the second term. None of the six terms of the
second and third row of \R{T4} have this
shape, but inserting the energy-momentum tensor into the last row of \R{T4} may generate such
terms. The first and the third term of the last row of \R{T4} become
\bey\nonumber
(\lambda^A u^A_l)_{,k}T^{Akl}\hspace{-.3cm} &=&\hspace{-.3cm}
\Big(\lambda^A{_{,k}}u^A_l +
\lambda^Au^A{_{l,k}}\Big)
\Big(\frac{1}{c^2}e^Au^{Ak}u^{Al}+
u^{Ak}p^{Al}+\frac{1}{c^2}q^{Ak}u^{Al}+t^{Akl}\Big)=
\\ \label{T9a3}
&=&\hspace{-.3cm} \lambda{^A}{_{,k}}u^{Ak}e^A
+\lambda^A u^A{_{l,k}} u^{Ak}p^{Al}
+\lambda^A{_{,k}}q^{Ak}
-\U{p^A\lambda^Au^{Ak}{_{,k}}}
+\lambda^Au^{A}{_{l,k}}\pi^{Akl},\hspace{.9cm}
\\ \label{Y14}
\lambda^A{_{l,k}}T^{Akl}\hspace{-.3cm}&=&\hspace{-.3cm}\lambda^A{_{l,k}} \Big(\frac{1}{c^2}e^Au^{Ak}u^{Al}+
u^{Ak}p^{Al}+\frac{1}{c^2}q^{Ak}u^{Al}+t^{Akl}\Big)
\eey
Summing up \R{T9a3} and \R{Y14} results in\footnote{the signs
$\U{\boxdot}$,
$\overbrace{\boxdot}$,
$\underbrace{\boxdot}$
and $\widetilde{\boxdot}$
mark terms which are related to each other in the sequel}
\bey\nonumber
(\lambda^A u^A_l)_{,k}T^{Akl}+\lambda^A{_{l,k}}T^{Akl}\ 
=\ \lambda{^A}{_{,k}}\Big(q^{Ak}+e^Au^{Ak}\Big)
+\ \lambda^Au^A{_{l,k}}\Big(\pi^{Akl}+u^{Ak}p^{Al}\Big)-
\\ \label{Y15}
-\U{p^A\lambda^Au^{Ak}{_{,k}}}
+\ \lambda^A{_{l,k}} \Big(\frac{1}{c^2}e^Au^{Ak}u^{Al}
+u^{Ak}p^{Al}+\frac{1}{c^2}q^{Ak}u^{Al}+t^{Akl}\Big).\hspace{.3cm}
\vspace{.3cm}\eey
Evidently, the term $-\U{p^A\lambda^Au^{Ak}{_{,k}}}$ belongs to the first term of \R{T4}.
After having performed the derivation and inserted the underlined term of \R{Y15}, 
the first term of \R{T4} becomes\footnote{\ $^\td$ is the "component time derivative" $\st{\td}{\boxplus}{^A}:=\boxplus^A{_{,k}}u^{Ak}$}
\bey\nonumber
&&u^{Ak}{_{,k}}\Big(s^A  -\kappa^A\rh^A-\lambda^A e^A - \U{p^A\lambda^A}
- \lambda^A_lp^{Al}\Big)+
\\ \nonumber
&&+\Big(s^A-\kappa^A\rh^A -\lambda^A e^A - \underbrace{p^A\lambda^A}
- \lambda^A_lp^{Al}\Big){^\td}
+\underbrace{(p^A\lambda^A){^\td}}=
\\ \label{T10}
&&=\ 
\Big[u^{Ak}\Big(s^A-\kappa^A\rh^A -\lambda^A e^A
-p^A\lambda^A
- \lambda^A_lp^{Al}\Big)\Big]_{,k}+\widetilde{(p^A\lambda^A){^\td}}.
\vspace{.3cm}\eey
Rearranging the entropy identity \R{T4} results in
\bey\nonumber
S^{Ak}{_{,k}} &\equiv&
\Big[u^{Ak}\Big(s^A-\kappa^A\rh^A -\lambda^A e^A
-p^A\lambda^A
-\lambda^A_lp^{Al}\Big)\Big]_{,k}+\widetilde{(p^A\lambda^A){^\td}} + \hspace{.5cm}
\\ \nonumber
&&+\ \Big[\ s^{Ak}
-\lambda^Aq^{Ak}
- \lambda^A_lt^{Akl}\Big]_{,k}+
\\ \nonumber
&&+\ \kappa^A\Gamma^A+\kappa^A{_{,k}}N^{Ak}
\\ \nonumber
&&+\ \lambda{^A}{_{,k}}\Big(q^{Ak}+e^Au^{Ak}\Big)
+\ \lambda^Au^A{_{l,k}}\Big(\pi^{Akl}+u^{Ak}p^{Al}\Big)+
\\ \nonumber
&&+\ \lambda^A{_{l,k}} \Big(\frac{1}{c^2}e^Au^{Ak}u^{Al}
+u^{Ak}p^{Al}+\frac{1}{c^2}q^{Ak}u^{Al}+t^{Akl}\Big)+
\\ \label{T10y}
&&+\ \lambda^A u^A_lk^{Akl}
+ \lambda^A_{l}k^{Akl}\ =\
\sigma^A + \varphi^A .
\vspace{.3cm}\eey
This entropy identity is incomplete: the multi-heat relaxation is missing which is gene\-rated by the
different partial temperatures of the components of the mixture. Because of lucidity, the
treatment of multi-heat relaxation is postponed and will be considered below in sect.\ref{PT}.
In the next section, we now specify $s^A$, $s^{Ak}$,  $\varphi^A$ and $\sigma^A$.

\subsection{Exploitation of the entropy identity\label{EXEI}}
\subsubsection{Entropy density, Gibbs and Gibbs-Duhem equations}

We now define the entropy rest density $s^A$ according to the first round bracket in \R{T10y}
\bey\nonumber
\mbox{\sf Setting III:}\hspace{8cm}
\\ \label{T10a}
s^A\ \st{\td}{=}\ \kappa^A\rh^A +\lambda^A e^A+p^A\lambda^A
+\lambda^A_lp^{Al},\hspace{1cm}
\eey
resulting in the specific rest entropy
\bee{T10a1}
\frac{s^A}{\rh}\ =\ \kappa^A\frac{\rh^A}{\rh}
+\lambda^A\frac{e^A}{\rh}
+p^A\lambda^A\frac{1}{\rh}
+\lambda^A_l\frac{p^{Al}}{\rh}.
\ee
A non-equilibrium state space --which is spanned by the independent variables--
contains the equilibrium variables $\rh^A$, $\rh$ and $e^A$ and beyond them
the non-equilibrium variables $p^{Al}$
extending the equilibrium sub-space in the sense of Extended Thermodynamics\footnote{If the energy-momentum tensor is presupposed to be symmetric --consequently
$p^{Al}=(1/c^2)q^{Al}$ is valid according to \R{J3c}-- the momentum density is
replaced by energy flux density which in  non-relativistic Extended Thermodynamics is set as a non-equilibrium variable, even if the stress tensor is non-symmetric.}
\C{JOU,MUE}. Consequently, we choose the state space
\bee{Y2}
{\sf z}^A\ =\ \Big(c^A,\frac{1}{\rh},\frac{e^A}{\rh},\frac{p^{Al}}{\rh}\Big),\qquad
c^A\ :=\ \frac{\rh^A}{\rh}
\ee
The corresponding Gibbs equation according to \R{T10a1} and \R{Y2} is
\bee{Y3}
\Big(\frac{s^A}{\rh}\Big)^\td\ =\
\kappa^A\st{\td}{c}{^A}
+\lambda^A \Big(\frac{e^A}{\rh}\Big)^\td
+p^A\lambda^A\Big(\frac{1}{\rh}\Big)^\td
+\lambda^A_l\Big(\frac{p^{Al}}{\rh}\Big)^\td.
\ee
Differentiation of \R{T10a1} results in the Gibbs-Duhem equation by taking
\R{Y3} into account
\bee{Y4}
0\ =\ \st{\td}{\kappa}\!{^A}c^A
+\st{\td}{\lambda}\!{^A}\frac{e^A}{\rh}
+(p^A\lambda^A)^\td\frac{1}{\rh}
+\st{\td}{\lambda}{^A_l}\frac{p^{Al}}{\rh},
\ee
resulting in
\bee{Y4a}
\widetilde{(p^A\lambda^A)^\td}\ =\
- \st{\td}{\kappa}\!{^A} \rh^A
-\st{\td}{\lambda}\!{^A} e^A
-\st{\td}{\lambda}{^A_l}p^{Al}.
\ee
Taking \R{Y4a} and \R{T10a} into account, the entropy identity \R{T10y} becomes
\bey\nonumber
S^{Ak}{_{,k}}\ \equiv\hspace{-.5cm} 
&&-\ub{\st{\td}{\kappa}\!{^A}\rh^A}
-\U{\st{\td}{\lambda}\!{^A} e^A}
-\widehat{\st{\td}{\lambda}{^A_l}p^{Al}} +
\\ \nonumber
&&+\ \Big[\ s^{Ak}
-\lambda^Aq^{Ak}
- \lambda^A_lt^{Akl}\Big]_{,k}+
\\ \nonumber
&&+\ \kappa^A\Big({^{(ex)}}\Gamma^A + {^{(in)}}\Gamma^A\Big)
+\ub{\kappa^A{_{,k}}N^{Ak}}+
\\ \nonumber
&&+\ \U{\lambda{^A}{_{,k}}}\Big(q^{Ak}+\U{e^Au^{Ak}}\Big)
+\ \lambda^Au^A{_{l,k}}\Big(\pi^{Akl}+u^{Ak}p^{Al}\Big)+
\\ \nonumber
&&+\ \widehat{\lambda^A{_{l,k}}} \Big(\frac{1}{c^2}e^Au^{Ak}u^{Al}
+\widehat{u^{Ak}p^{Al}}+\frac{1}{c^2}q^{Ak}u^{Al}+t^{Akl}\Big)+
\\ \label{Y15z}
&&+\lambda^A u^A_lk^{Al}
+ \lambda^A_{l}k^{Al}\ 
=\ \sigma^A + \varphi^A .
\eey
The marked terms cancel each other. Taking \R{K8c}$_2$ and \R{L3}$_1$ into account, we consider
\bey\nonumber
0 &=& -\ub{\st{\td}{\kappa}\!{^A}\rh^A}+\ub{\kappa^A{_{,k}}N^{Ak}}\ =\
-\kappa^A{_{,k}}\Big(J^{Ak}-J^{Ak}\Big)\ =\ 
\\ \nonumber
&=&-\kappa^A{_{,k}}J^{Ak}
+\kappa^A{_{,k}}\Big(J^{Am}h^{Ak}_m + \rh^Aw^Au^{Ak}\Big)\ =\
\\ \label{Y15w}
&=&-\kappa^A{_{,k}}\Big(\ub{J^{Ak}-w^AN^{Ak}}_{J^{Am}h^{Ak}_m}\Big)
+\Big(\kappa^AJ^{Am}h^{Ak}_m\Big)_{,k}-\kappa^A\Big(J^{Am}h^{Ak}_m\Big)_{,k}
\vspace{.3cm}\eey
This zero contains the diffusion flux which does not appear up to here in the entropy
identity \R{aT2}. That means, the diffusion is missing in \R{aT2}, and we will not ignore
the underbraced terms in \R{Y15w}$_1$, but we insert \R{Y15w}$_3$ into \R{Y15z}.
Consequently, the entropy identity results in
\bey\nonumber
S^{Ak}{_{,k}}\ \equiv\hspace{-.5cm} 
&& \Big[\ s^{Ak}
-\lambda^Aq^{Ak}
- \lambda^A_lt^{Akl}+\kappa^AJ^{Am}h^{Ak}_m\Big]_{,k}+
\\ \nonumber
&&+\ \kappa^A\Big(\U{{^{(ex)}}\Gamma^A} + {^{(in)}}\Gamma^A
-(J^{Am}h^{Ak}_m)_{,k}\Big)
-\kappa^A{_{,k}}J^{Am}h^{Ak}_m +
\\ \nonumber
&&+\ \lambda{^A}{_{,k}}q^{Ak}
+\ \lambda^Au^A{_{l,k}}\Big(\pi^{Akl}+u^{Ak}p^{Al}\Big)+
\\ \nonumber
&&+\ \lambda^A{_{l,k}} \Big(\frac{1}{c^2}e^Au^{Ak}u^{Al}
+\frac{1}{c^2}q^{Ak}u^{Al}+t^{Akl}\Big)+
\\ \label{Y15v}
&&+\U{\lambda^A u^A_lk^{Akl}}
+\U{\lambda^A_{l}k^{Akl}}\ 
=\ \sigma^A + \varphi^A .
\eey
The underlined terms belong to the interaction of the system with its environment.
We now specify the entropy flux density $s^{Ak}$ and the entropy supply $\varphi^A$
in the next section.

\subsubsection{Entropy flux, --supply and --production}

According to the first row of \R{Y15v}, we define the entropy flux density
\bey\nonumber
\mbox{\sf Setting IV:}\hspace{8cm}
\\ \label{T7} 
s^{Ak}\ \st{\td}{=}\ \lambda^Aq^{Ak}
+\lambda^A_lt^{Akl}
-\kappa^AJ^{Am}h^{Ak}_m.
\hspace{1.3cm}
\vspace{.3cm}\eey
We now split the entropy identity \R{Y15z} into the entropy production and the entropy
supply.  For this end, we need a criterion to distinguish between entropy production and supply.
Such a criterion is clear for discrete systems: a local isolation suppresses the entropy
supply but not the entropy production. Isolation means: the three underlined terms in \R{Y15v}
vanish, if the $^A$-component is isolated from the exterior of the mixture. Consequently, we define
the entropy supply as follows
\bey\nonumber
\mbox{\sf Setting V:}\hspace{8cm}
\\ \label{T8}
\varphi^A\ \st{\td}{=}\
\kappa^A {^{(ex)}}\Gamma^A
+\lambda^A u_l^Ak^{Al} + \lambda^A_mk^{Am}\hspace{1.3cm}
\eey
with the result that the entropy identity \R{Y15v} transfers into the entropy production density by
taking \R{T7} and \R{T8} into account
\bey\nonumber
\sigma^A\ =\ \hspace{-.5cm}
&&\ \kappa^A\Big({^{(in)}}\Gamma^A
-(J^{Am}h^{Ak}_m)_{,k}\Big)
-\kappa^A{_{,k}}J^{Am}h^{Ak}_m +
\\ \nonumber
&&+\ \lambda{^A}{_{,k}}q^{Ak}
+\ \lambda^Au^A{_{l,k}}\Big(\pi^{Akl}+u^{Ak}p^{Al}\Big)+
\\ \label{T9}
&&+\ \lambda^A{_{l,k}} \Big(\frac{1}{c^2}e^Au^{Ak}u^{Al}
+\frac{1}{c^2}q^{Ak}u^{Al}+t^{Akl}\Big).
\vspace{.3cm}\eey
As expected, the entropy production is composed of
terms which are a product of "forces" and "fluxes" as in the non-relativistic
case\footnote{The mass production $^{(in)}\Gamma^A$ due to chemical reactions can be expressed by the time rate of the reaction velocity, see \R{Z13c} in sect.\ref{SE}.}.
The expressions $s^A,\ s^{Ak},\ \varphi^A$ and $\sigma^A$  
contain accessory variables which are introduced for formulating the entropy identity \R{aT2}
playing up to here the role of place-holders. Their physical meaning is discussed in the next section.

\subsection{Accessory variables\label{ACVA}}

Starting out with \R{T10a}, we have the following equation of physical dimensions
\bee{Z1}
[s^{A}]\ =\ [\lambda^A][e^A].
\ee
Taking \R{T2w}$_1$ and \R{J6} into account, we obtain
\bee{Z2}
\frac{N}{m^2}\frac{1}{K}\ =\ [\lambda^A] \frac{N}{m^2}
\quad\longrightarrow\quad
[\lambda^A]\ =\ \frac{1}{K},
\ee
that means, $\lambda^A$ is a reciprocal temperature belonging to the
$^A$-component. Therefore, we accept the following
\bey\nonumber
\mbox{\sf Setting VI:}\hspace{6cm}
\\ \label{Z3}
\lambda^A\ \st{\td}{=}\ \frac{\nu^A}{\Theta^A},\hspace{3cm}
\eey
with the partial temperature  $\Theta^A$ of the $^A$-component\footnote{This
temperature is a non-equilibrium one, the contact temperature \C{MUTEMP,MUTEMP1} which should not be confused with the thermostatic equilibrium temperature $\Theta^A_{eq}=T,\ \wedge A$.} and a scalar $\nu^A$ which is suitably chosen below..
\vspace{.3cm}\newline
Starting out with \R{T7} , we have the following equation of physical dimensions
\bee{Z4}
[s^{Ak}]\ =\ [\kappa^A][J^{Am}][h_m^{Ak}].
\ee
Taking \R{T2w}$_2$, \R{K8c}$_2$ and \R{J5}$_1$ into account, we obtain 
\bee{Z5}
\frac{N}{ms}\frac{1}{K}\ =\ [\kappa^A]\frac{kg}{m^3}\frac{m}{s}1
\quad\longrightarrow\quad
[\kappa^A]\ =\ \frac{m^2}{s^2}\frac{1}{K}.
\ee
We know from the non-relativistic Gibbs equation that the chemical potentials $\mu^A$
have the physical dimension of the specific energy $e^A/\rh^A$
\bee{Z5a}
[\mu^A]\ = \frac{[e^A]}{[\rh^A]}\ =\ \frac{N}{m^2}\frac{m^3}{kg}\ =\ K[\kappa^A].
\ee
Consequently, we make the following choice by taking \R{Z5a} into consideration
\bey\nonumber
\mbox{\sf Setting VII:}\hspace{6cm}
\\ \label{Z7}
\kappa^A\ \st{\td}{=}\ \frac{\mu^A}{\Theta^A}.\hspace{3cm}
\vspace{.3cm}\eey
Starting out with \R{T7} and \R{J6}$_1$, we have the following equation of physical dimensions
\bee{Z8}
[s^{Ak}]\ =\ [\lambda^{Ak}][t^{Akl}]\ =\ [\lambda^{Ak}][p^A].
\ee
Taking \R{T2w}$_2$ and \R{J7} into account, we obtain
\bee{Z9}
\frac{N}{ms}\frac{1}{K}\ =\ [\lambda^{Ak}]\frac{N}{m^2}
\quad\longrightarrow\quad
[\lambda^{Ak}]\ =\ \frac{m}{s}\frac{1}{K},
\ee
that means, $\lambda^{Ak}$ is proportional to a velocity and at the same time perpendicular
to $u^{Ak}$ according to \R{bT2}$_2$. Consequently, the velocity $u^m$ of the mixture remains
for defining $\lambda^{Ak}$
\bey\nonumber
\mbox{\sf Setting VIII:}\hspace{6cm}
\\ \label{Z10}
\lambda^{Ak}\ \st{\td}{=}\ \frac{1}{\Theta^A}u^mh^{Ak}_m.\hspace{3cm}
\vspace{.3cm}\eey
Inserting the accessory variables into the expression of entropy density \R{T10a}, of entropy flux
density \R{T7} and of entropy supply \R{T8}, we obtain
\byy{Z14}
s^A &=& \frac{1}{\Theta^A}\Big(\mu^A\rh^A + \nu^A(e^A+p^A)
+u_mp^{Am}\Big),
\\ \label{Z15}
s^{Ak} &=& \frac{1}{\Theta^A}\Big(\nu^Aq^{Ak}
-\mu^AJ^{Am}h^{Ak}_m
+u_mt^{Akm}\Big),
\\ \label{Z16}
\varphi^A &=& \frac{1}{\Theta^A}\Big(\mu^A{^{(ex)}}\Gamma^A+
\nu^A u_l^Ak^{Al} + u_ph^{Ap}_mk^{Am}\Big).\hspace{.6cm}
\eey
The entropy production density \R{T9} results by use of \R{Z7}, \R{Z10} and \R{L3}$_2$
\bey\nonumber
\sigma^A\ =\ \hspace{-.5cm}
&&\ \frac{\mu^A}{\Theta^A}\Big({^{(in)}}\Gamma^A
-(J^{Am}h^{Ak}_m)_{,k}\Big)
-\Big(\frac{\mu^A}{\Theta^A}\Big)_{,k}J^{Am}h^{Ak}_m +
\\ \nonumber
&&+\ \Big(\frac{\nu^A}{\Theta^A}\Big)_{,k}q^{Ak}
+\ \frac{\nu^A}{\Theta^A}u^A{_{l,k}}\Big(\pi^{Akl}
+u^{Ak}p^{Al}\Big)+
\\ \label{T9z}
&&+\ \Big(\frac{1}{\Theta^A}u^mh^A_{ml}\Big)_{,k}
\Big(\frac{1}{c^2}e^Au^{Ak}u^{Al}
+\frac{1}{c^2}q^{Ak}u^{Al}+t^{Akl}\Big).
\vspace{.3cm}\eey
The first four terms of the entropy production describe the four classical reasons of irreversibility:
chemical reactions, diffusion,  heat conduction and internal friction with a modified non-equilibrium
viscous tensor. The last term of \R{T9z}\footnote{which vanishes in equilibrium
and for free 1-component systems, as we will see below} is typical for an interacting
$^A$-component as a part of the mixture according to its LHS factor.
\vspace{.3cm}\newline
Up to now, a further phenomenon of irreversibility was not taken into consideration: the multi-heat relaxation which is discussed in the next section.

\subsection{Multi-heat relaxation and the partial temperatures\label{PT}}

Because the different components of the mixture have different partial (reciprocal) temperatures
$\lambda^A,\ A=1,2,...,Z$, a multi-heat relaxation\footnote{do not take multi-heat
relaxation for the heat conduction which is caused by temperature
gradients} takes place which is an irreversible phenomenon. Consequently, multi-heat relaxation
has to be taken into account in the entropy identity by adding a suitable zero as done in \R{aT2}.
\vspace{.3cm}\newline
A heat transfer $H^{ABk}$ between two components of the mixture --$A$ and $B$--
takes place by multi-heat relaxation, if the corresponding temperatures of the components are different from each other. Consequently, the entropy exchange
between these two components is 
\bey\nonumber
\mbox{\sf Setting IX:}\hspace{8cm}
\\ \label{A10}
G^{ABk}\ :=\ H^{ABk}\Big(\frac{1}{\Theta^A}-\frac{1}{\Theta^B}\Big), 
\quad H^{BBk}\ \equiv\ 0.
\eey
As the energy flux density \R{J3}$_1$ and \R{J8}, the multi-heat transfer satisfies
\byy{A11}
H^{ABk}u^A_k\ =\ 0,\quad H^{ABk}u^B_k\ =\ 0,\quad H^{ABk}\ =\ -H^{BAk},
\\ \label{A12}
[H^{ABk}]\ =\ [q^{AK}]\ =\ [e ^A]\frac{m}{s},\quad [G^{ABk}]\ =\ 
[e ^A]\frac{m}{s}\frac{1}{K}\ =\ [s^{Ak}].
\vspace{.3cm}\eey
For the $^A$-component, this results according to \R{A11}$_3$ in
\byy{A13}
H^{Ak} &:=& \sum_B H^{ABk},\quad \sum_{AB} H^{ABk}\ =\ 0,
\\ \nonumber
G^{Ak} &:=& \sum_B H^{ABk}\Big(\frac{1}{\Theta^A}-\frac{1}{\Theta^B}\Big)\ =\
\\ \label{A14}
&=& H^{Ak}\frac{1}{\Theta^A}-\sum_B H^{ABk}\frac{1}{\Theta^B},\quad G^{Ak}u^A_k\ =\ 0,
\\ \label{A15}
\sum_A G^{Ak} &=& \sum_{AB} H^{ABk}\Big(\frac{1}{\Theta^A}-\frac{1}{\Theta^B}\Big)\ \neq\ 0,\ \mbox{if}\
\Theta^A\neq\Theta^B.
\vspace{.3cm}\eey
The entropy exchange of the $^A$-component according to multi-heat exchange \R{A14}$_1$
has now to be introduced into the entropy identity \R{aT2}. Because $G^{Ak}$ has the same
physical dimension as $s^{Ak}$, the zero
\bey\nonumber
0 &=&\Big(G^{Ak} - \sum_B H^{ABk}\Big(\frac{1}{\Theta^A}-\frac{1}{\Theta^B}\Big)\Big)_{,k}\ =
\\ \label{A16}
&=&G^{Ak}{_{,k}} - \sum_B H^{ABk}{_{,k}}\Big(\frac{1}{\Theta^A}-\frac{1}{\Theta^B}\Big) - \sum_B H^{ABk}\Big(\frac{1}{\Theta^A}-\frac{1}{\Theta^B}\Big)_{,k}
\eey
can be directly introduced into the entropy identity without defining an additional accessory variable.
According to \R{T4}, the three terms of \R{A16} are attached as follows
\byy{A16x}
+\sum_B H^{ABk}\Big(\frac{1}{\Theta^A}-\frac{1}{\Theta^B}\Big)_{,k}
&\longrightarrow&\mbox{entropy production density},
\\ \label{A17x}
+\sum_B H^{ABk}{_{,k}}\Big(\frac{1}{\Theta^A}-\frac{1}{\Theta^B}\Big)
&\longrightarrow&\mbox{entropy supply},
\\ \label{A18x}
-G^{Ak}{_{,k}}
&\longrightarrow&\mbox{entropy exchange density}.
\vspace{.3cm}\eey
Introducing these terms as drawn in sect.\ref{IC1}, \R{Z14} to \R{T9z} yield
\byy{Z14p}
s^A &=& \frac{1}{\Theta^A}\Big(\mu^A\rh^A + \nu^A(e^A+p^A)
+u_mp^{Am}\Big),
\\ \label{Z15p}
s^{Ak} &=& \frac{1}{\Theta^A}\Big(\nu^Aq^{Ak}
-\mu^AJ^{Am}h^{Ak}_m
+u_mt^{Akm}\Big)+G^{Ak},
\\ \label{Z16p}
\varphi^A &=& \frac{1}{\Theta^A}\Big(\mu^A{^{(ex)}}\Gamma^A+
\nu^A u_l^Ak^{Al} + u_ph^{Ap}_mk^{Am}\Big)+\sum_B H^{ABk}{_{,k}}\Big(\frac{1}{\Theta^A}-\frac{1}{\Theta^B}\Big),
\\ \nonumber
\sigma^A &=& 
\frac{\mu^A}{\Theta^A}\Big({^{(in)}}\Gamma^A
-(J^{Am}h^{Ak}_m)_{,k}\Big)
-\Big(\frac{\mu^A}{\Theta^A}\Big)_{,k}J^{Am}h^{Ak}_m +
\\ \nonumber
&&+\ \Big(\frac{\nu^A}{\Theta^A}\Big)_{,k}q^{Ak}
+\ \frac{\nu^A}{\Theta^A}u^A{_{l,k}}\Big(\pi^{Akl}
+u^{Ak}p^{Al}\Big)
+\sum_B H^{ABk}\Big(\frac{1}{\Theta^A}-\frac{1}{\Theta^B}\Big)_{,k}+
\hspace{.7cm}
\\ \label{T9zp}
&&+\ \Big(\frac{1}{\Theta^A}u^mh^A_{ml}\Big)_{,k}
\Big(\frac{1}{c^2}e^Au^{Ak}u^{Al}
+\frac{1}{c^2}q^{Ak}u^{Al}+t^{Akl}\Big).
\vspace{.3cm}\eey
The six terms of the entropy production density \R{T9zp} have the following meaning:
\begin{flushleft}$\bullet\quad$
diffusion modified chemical reaction:\ $(\mu^A/\Theta^A)\Big({^{(in)}}\Gamma^A
+(J^{Am}h^{Ak}_m)_{,k}\Big)$\\
$\bullet\quad$diffusion:\ $(\mu^A/\Theta^A){_{,k}}J^{Am}h^{Ak_m}$,\\
$\bullet\quad$heat conduction:\ $(\nu^A/\Theta^A){_{,k}}q^{Ak}$,\\
$\bullet\quad$multi-component modified internal friction:\ $(\nu^A/\Theta^A)u^A{_{l,k}}\Big(\pi^{Akl}+u^{Ak}p^{Al}\Big)$,\\
$\bullet\quad$multi-heat relaxation:\ $\sum_B H^{ABk}\Big((1/\Theta^A)-(1/\Theta^B)\Big)_{,k}$,\\
$\bullet\quad$multi-component interaction\footnote{This term vanishes in equilibrium and for 1-component systems: 
see sect.\ref{FC}}:\ 
$(u^mh^A_{ml}/\Theta^A){_{,k}} \Big(\frac{1}{c^2}e^Au^{Ak}u^{Al}
+\frac{1}{c^2}q^{Ak}u^{Al}+t^{Akl}\Big)$.
\end{flushleft}

\subsection{The 4-entropy}

We need the 4-entropy of the $^A$-component for describing thermodynamics of a mixture.
Starting out with \R{T2}, \R{Z14}, \R{A18x} and \R{Z15}, we obtain 
\bey\nonumber
S^{Ak} &=&
\K{A}u^{Ak}\Big\{
\mu^A\rh^A + \nu^A \Big(e^A+p^A\Big)+ u_mp^{Am}+
\\ \nonumber
&&+\K{A}\Big\{\nu^A q^{Ak}
-\mu^AJ^{Am}h^{Ak}_m
+ u_mt^{Akm}\Big\}+G^{Ak}\ =\
\\ \nonumber
&=&
\K{A}\Big\{\nu^A\Big(e^Au^{Ak}+q^{Ak}\Big)
+u_m\Big(u^{Ak}p^{Am}+t^{Akm}\Big)\Big\}+
\\ \label{E1}
&&+\frac{\mu^A}{\Theta^A}N^{Ak}+\frac{\nu^A}{\Theta^A}p^Au^{Ak}
-\frac{\mu^A}{\Theta^A}J^{Am}h^{Ak}_m+G^{Ak}.
\eey
Inserting \R{K11b} and \R{L3}$_2$, \R{E1} results in
\bee{E2}
S^{Ak}\ =\
\K{A}\Big\{\nu^AQ^{Ak}+u_m\tau^{Akm}\Big\}
+\frac{\mu^A}{\Theta^A}\Big((1+w^A)N^{Ak}-J^{Ak}\Big)
+\frac{\nu^A}{\Theta^A}p^Au^{Ak}+G^{Ak}.
\vspace{.3cm}\ee
The transition from the interacting $^A$-component to the free 1-component system is considered
in sect.\ref{FC} and that to the mixture in sect.\ref{TM}. All quantities introduced up to here are non-equilibrium ones, because we did not
consider equilibrium conditions up to now. This will be done in the next section.

\subsection{Equilibrium\label{EQ}}
\subsubsection{Equilibrium conditions}

Equilibrium is defined by {\em equilibrium conditions} which are divided into
{\em basic} and {\em supplementary} ones \C{MUBO,MUBO1}. The basic equilibrium
conditions are
given by vanishing entropy production, vanishing entropy flux density and vanishing
entropy supply\footnote{The sign $\doteq$ stands for a setting which implements an
equilibrium condition.}:
\bee{P23} 
\sigma^{A}_{eq}\ \doteq\ 0\quad\wedge\quad s^{Ak}_{eq}\ \doteq\ 0\quad\wedge\quad
\varphi^A_{eq}\ \doteq\ 0.
\ee
A first supplementary equilibrium condition is the vanishing of all diffusion flux densities.
According to \R{K8c}$_1$, we obtain 
\bee{P23a}
J^{Aeq}_k\ \doteq\ 0\quad\longrightarrow\quad
u^{Aeq}_k\ =\ f^A_{eq}u^{eq}_k\quad\longrightarrow\quad
c^2\ =\ \ f^A_{eq}u^{eq}_k u^{Ak}_{eq}.
\ee
Taking \R{K6b}$_1$ into account, \R{P23a}$_3$ results in
\bee{P23b}
(f^A_{eq})^2\ =\ 1\quad\longrightarrow\quad f^A_{eq}\ =\ \pm 1.
\ee
Consequently, we have to demand beyond \R{P23a}$_1$ the supplementary equilibrium
condition that the mass densities are additive in equilibrium. We obtain according to \R{K6b}$_2$
and \R{K9a}$_2$
\bee{P23c}
\rh_{eq}\ \doteq\ \sum_A\rh^A_{eq}\ \longrightarrow\ f^A_{eq}\ =\ 1\ \longrightarrow\
w^A_{eq}\ =\ 0.
\ee
Taking \R{P23a}$_2$ and \R{Z10} into account, \R{P23c}$_2$ yields
\bee{P23d}
u^{Aeq}_k\ =\ u^{eq}_k\ \longrightarrow\
{\lambda}{^{Ak}_{eq}}\ =\ 0,\quad g^{Am}_{eq}\ =\ 0.
\vspace{.3cm}\ee
Further supplementary equilibrium conditions are given by vanishing covariant
time derivatives, except that of the four-velocity:
\bee{P24}
\boxplus^\bullet_{eq}\ \doteq\ 0,\qquad\boxplus\
\neq\ u^l,
\ee
that means $\st{\td}{u}\!{^l_{eq}}$ is in general not zero in equilibrium. Consequently,
the time derivatives of all expressions which contain the 4-velocity must be calculated
separately, as we will see below.  
\vspace{.3cm}\newline
According to \R{P24}$_1$, we obtain
\bee{P25}
\st{\td}{\rh}\!{^A_{eq}}\ =\ 0,\qquad
\Big(\frac{\nu^A}{\Theta^A}\Big)^\td_{eq}\ =\ 0,
\ee
and the (3+1)-components of the energy-momentum tensor, \R{J2} and \R{J3},
satisfy
\bee{aP25}
\st{\td}{e}{^A_{eq}}=0,\quad \st{\td}{p}{^{Al}_{eq}}=0,
\quad \st{\td}{q}{^{Ak}_{eq}}=0,\quad \st{\td}{p}{^A_{eq}}=0,
\quad \st{\td}{\pi}{^{Akl}_{eq}}=0.
\vspace{.3cm}\ee
Starting out with \R{K6b}$_1$, we  have
\bee{P25b}
\st{\td}{f}\!{^A_{eq}}\ =\ \frac{1}{c^2}\Big(\st{\td}{u}{^{Aeq}_m}u^m_{eq}+
{u}{^{Aeq}_m}\st{\td}{u}\!{^m_{eq}}\Big).
\ee
Taking \R{P23d}$_1$ into account, this results in
\bee{bP25}
\st{\td}{f}\!{^A_{eq}}\ =\ 0\quad\longrightarrow\quad
\st{\td}{w}\!{^A_{eq}}\ =\ 0. 
\vspace{.3cm}\ee
In equilibrium, we have according to \R{P23d}$_1$ and \R{K15}
\bee{Y16d}
h^{Am}_{leq}\ =\ h^{m}_{leq},
\ee
according to \R{Z10} resulting in
\bee{Y16c}
\lambda^{Aeq}{_{l,k}}\ =\ 
\Big(\frac{1}{\Theta^A}u_mh^{Am}_l\Big)_{,k}^{eq}\ =\ 0. 
\ee
Consequently, the time derivatives of the accessory variables vanish in equilibrium,
and according to \R{Y3} and \R{Y4}, Gibbs and Gibbs--Duhem equations
are identically satisfied in equilibrium.
\vspace{.3cm}\newline
Another supplementary equilibrium condition is the vanishing of the mass production terms in
\R{K7a}$_{3,4}$
\bee{Y15a}
{^{(ex)}}\Gamma^A_{eq}\ \doteq\ 0\ \wedge\ {^{(in)}}\Gamma^A_{eq}\ \doteq\ 0
\ \longrightarrow\  \Gamma^A_{eq}\ =\ 0
\ee
Thus, we obtain from \R{K2}, \R{P25}$_1$ and \R{Y15a} 
\bee{Y16}
\varrho^A{_{,k}}u^{Ak}+ \varrho^Au^{Ak}{_{,k}}\ =\ \Gamma^A
\quad\longrightarrow\quad u^{Ak}_{eq}{_{,k}}\ =\ 0.
\vspace{.3cm}\ee
Taking \R{P23c}$_3$, \R{P23d}$_1$  and \R{J2}$_2$ into account, the entropy density \R{Z14p} becomes in equilibrium
\bee{W1}
s^A_{eq}\ =\ \frac{1}{\Theta^A_{eq}}\Big(\mu^A_{eq}\rh^A _{eq}
+\nu^A_{eq}(e^A_{eq} + p^A_{eq})\Big).
\ee
Except of $\nu^A_{eq}$,
this is the usual expression for the entropy density in thermostatics\footnote{Below, we will see that $\nu^A_{eq}=1$}. The energy density and the pressure are here defined by
the (3+1)-decomposition \R{J1} of the energy-momentum tensor.
The chemical potential is as well as the temperature introduced as an accessory variable.
\vspace{.3cm}\newline
The equilibrium temperature in \R{W1} is characterized by vanishing multi-heat relaxation
\bee{W1a}
\Theta^A_{eq}\ \doteq\ \Theta^B_{eq}\ \doteq\ \Theta^C_{eq}\ \doteq...\ =:\ \Theta_{eq}.
\ee
Often one can find in literature \C{KUI} the case of equilibrium of multi-heat relaxation:
although out of equilibrium, only one temperature is considered in multi-component systems. This case is realistic, if the relaxation of multi-heat relaxation to equilibrium is
remarkably faster than that of the other non-equilibrium variables \C{MUIV}. As demonstrated, the non-equilibrium entropy density
\R{Z14p} does not depend on multi-heat relaxation, a reason why often equilibrium of
multi-heat relaxation is presupposed without any remark. 
\vspace{.3cm}\newline
Taking \R{P23}$_2$, \R{P23a}$_1$, \R{P23d}$_1$  and \R{W1a} into account, the entropy flux density \R{Z15p} va\-nishes in equilibrium
\bee{Y16a}
0\ =\ q^{Aeq}_k,
\ee
and finally taking \R{Y15a}$_1$, \R{P23}$_3$, \R{P23c}$_3$, \R{P23a}$_1$ and \R{P23d}$_1$
into account, the entropy supply \R{Z16p} results in
\bee{W2}
0\ =\ u^{Aeq}_lk^{Al}_{eq},
\ee
that means, the power exchange vanishes in equilibrium.
\vspace{.3cm}\newline
The entropy production \R{T9zp} has to vanish in equilibrium according to the basic
equi\-li\-brium condition \R{P23}$_1$. Taking \R{Y15a}$_2$, \R{P23a}$_1$, 
\R{Y16a}, \R{aP25}$_2$, \R{P23d}$_1$ and \R{W1a} into account and using \R{P24},
\R{T9zp} results in
\bee{Y16b}
0\ =\ u^{Aeq}{_{l,k}}\pi^{Akl}_{eq}.
\vspace{.3cm}\ee
As demonstrated, equilibrium of an $^A$-component in the mixture is described by three basic
equilibrium conditions \R{P23} and six supplementary ones: \R{P23a}$_1$, \R{P23c}$_1$,
\R{P24}, \R{Y15a}$_{1,2}$ and \R{W1a}. Often, we can find in literature \C{M85,EU}
equilibrium conditions
which are different from those postulated here. The reason for that is, that entropy production and
supply and the entropy flux as starting-points for the basic equilibrium conditions differ from the
expressions \R{Z14p} to \R{T9zp}. Such different equilibrium conditions and their derivations are
considered in the next two sections.

\subsubsection{Killing relation of the 4-temperature}

Starting out with \R{T9a3}$_1$, we now consider the following relations
\byy{B-1}
(\lambda^A{_{,k}}u^A_l+\lambda^A u^A_{l,k})\frac{1}{c^2}e^Au^{Ak}u^{Al}
&=& \st{\td}{\lambda}\!{^A}e^A,
\\ \label{B-2}
(\lambda^A{_{,k}}u^A_l+\lambda^A u^A_{l,k})u^{Ak}p^{Al}
&=& -\lambda^A u^A_l\st{\td}{p}\!{^{Al}},
\\ \label{B-3}
(\lambda^A{_{,k}}u^A_l+\lambda^A u^A_{l,k})\frac{1}{c^2}q^{Ak}u^{Al}
&=& \lambda^A{_{,k}}q^{Ak},
\\ \label{B-4}
-(\lambda^A{_{,k}}u^A_l+\lambda^A u^A_{l,k})p^Ah^{Akl}
&=& -\lambda^A p^A  u^{Ak}{_{,k}},
\\ \label{B-5}
(\lambda^A{_{,k}}u^A_l+\lambda^A u^A_{l,k})\pi^{Akl}
&=& \lambda^A u^A_{l,k}\pi^{Akl}.
\eey
Taking \R{B-2}, \R{B-3} and \R{B-5} into account,  we obtain from \R{T9a3}$_1$
\bey\nonumber
(\lambda^A u^A_l)_{,k}
\Big(T^{Akl}-\frac{1}{c^2}e^Au^{Ak}u^{Al} +p^Ah^{Akl}\Big)\ =\hspace{4.5cm}
\\ \label{B1}
=\ -\lambda^Au^{A}_l\st{\td}{p}\!{^{Al}}
+\lambda^A{_{,k}}q^{Ak}
+\lambda^Au^A{_{l,k}}\pi^{Akl}.\hspace{1cm}
\eey
Replacing the second row of \R{T9zp} by \R{B1} yields for the entropy production of vanishing multi-heat relaxation
\bey\nonumber
\sigma^A &=& 
\ \lambda^A\mu^A\Big({^{(in)}}\Gamma^A-(J^{Am}h^{Ak}_m)_{,k}\Big)
-(\lambda^A\mu^A){_{,k}}J^{Am}h^{Ak}_m +
\\ \nonumber
&&+(\lambda^A u^A_l)_{,k}
\Big(T^{Akl}-\frac{1}{c^2}e^Au^{Ak}u^{Al} +p^Ah^{Akl}\Big)+
\\ \label{T7gy}
&&+\ \Big(\frac{1}{\Theta^A}u^mh^A_{ml}\Big)_{,k}\Big(\frac{1}{c^2}e^Au^{Ak}u^{Al}
+\frac{1}{c^2}q^{Ak}u^{Al}+t^{Akl}\Big).
\eey
Evident is that
\bee{B1a}
(\lambda^A u^A_l)_{,k}
\Big(T^{Akl}-\frac{1}{c^2}e^Au^{Ak}u^{Al} +p^Ah^{Akl}\Big)\ =\ 0
\ee
is not a sufficient condition for equilibrium because the equilibrium conditions
\R{Y15a}$_2$, \R{P23a}$_1$  and \R{P23d}$_1$ are not necessarily satisfied and the entropy production \R{T7gy} does not vanish.
\vspace{.3cm}\newline
 If the energy-momentum tensor is symmetric, \R{B1a} results in
\bee{B1b}
T^{Akl}=T^{Alk}:\quad
\Big[(\lambda^A u^A_l)_{,k}+(\lambda^A u^A_k)_{,l}\Big]
\Big(T^{Akl}-\frac{1}{c^2}e^Au^{Ak}u^{Al} +p^Ah^{Akl}\Big)\ =\ 0,
\ee
an expression which as well as \R{B1a} is not sufficient for equilibrium. Consequently, the
{\em Killing relation of the 4-temperature} $\lambda^A u^A_{l}$
\bee{V8}
\Big[(\lambda^A u^A_{l})_{,k} + (\lambda^A u^A_{k})_{,l}\Big]\ =\ 0
\ee 
is also not sufficient for equilibrium\footnote{a fact
which is well-known \C{MUBO1}}. If equilibrium is presupposed, the equilibrium conditions \R{Y15a}$_2$,
\R{P23a}$_1$  and \R{P23d}$_1$ are satisfied, the entropy production vanishes and
\byy{V9}
(\lambda^A u^A_l)_{,k}^{eq}
\Big(T^{Akl}-\frac{1}{c^2}e^Au^{Ak}u^{Al} +p^Ah^{Akl}\Big)^{eq}\ =\ 0,
\\ \label{V9a}
T^{Akl}_{eq}=T^{Alk}_{eq}:\quad
\Big[(\lambda^A u^A_l)_{,k}+(\lambda^A u^A_k)_{,l}\Big]^{eq}
\Big(T^{Akl}-\frac{1}{c^2}e^Au^{Ak}u^{Al} +p^Ah^{Akl}\Big)^{eq}\ =\ 0
\eey
are necessary conditions\footnote{but as discussed, not sufficient conditions} for equilibrium
according to \R{T7gy}. There are different possibilities to satisfy \R{V9} and \R{V9a} which are discussed in the next section.

\subsubsection{The gradient of the 4-temperature\label{G4T}}

The necessary condition for equilibrium \R{V9} can be differently satisfied generating
different types of equilibria. There are three possibilities:
\bey\nonumber
&&\hspace{-3.7cm}\mbox{If equilibrium exists, one of the following three conditions is valid:}  
\\ \label{V10}
(\lambda^A u^A_l)_{,k}^{eq} &=& 0\ \longrightarrow\
\lambda^{Aeq}_{,k}u^{Aeq}_l+\lambda^A_{eq} u^{Aeq}_{l,k}\ =\ 0,
\\ \label{V11}
T^{Akl}_{eq} &=& \frac{1}{c^2}e^A_{eq}u^{Ak}_{eq}u^{Al}_{eq} -p^A_{eq}h^{Akl}_{eq},
\\ \label{V12}
(\lambda^A u^A_l)^{eq}_{,k}\ \neq\ 0&\wedge&
\Big[T^{Akl}_{eq}\ \neq\ \frac{1}{c^2}e^A_{eq}u^{Ak}_{eq}u^{Al}_{eq}
-p^A_{eq}h^{Akl}_{eq}\Big],\quad\mbox{and \R{V9} is valid.}
\vspace{.3cm}\eey
Multiplication of \R{V10}$_2$ with $u^{Al}_{eq}$ results in
\bee{V13}
\lambda^{Aeq}_{,k}\ =\ 0\ \wedge\ u^{Aeq}_{l,k}\ =\ 0,
\ee
that means, \R{V10} represents an intensified equilibrium because additionally to the usual
equilibrium conditions mentioned in sect.\ref{EQ}, \R{V13} is valid. 
\vspace{.3cm}\newline
If \R{V11} is valid, the equilibrium exists in a perfect material whose entropy production is zero.
If the considered material is not perfect and if the equilibrium is not intensified, \R{V12} is valid,
and the question arises, whether \R{V9} can be valid under these constraints. To answer this question, we consider \R{B-1} to \R{B-5} in equilibrium. According to the equilibrium conditions, we obtain
\byy{B-6}
(\lambda^A_{,k}u^A_l+\lambda^A u^A_{l,k})^{eq}\frac{1}{c^2}e^Au^{Ak}_{eq}u^{Al}_{eq}\
&=& 0,
\\ \label{B-7}
(\lambda^A_{,k}u^A_l+\lambda^A u^A_{l,k})^{eq}u^{Ak}_{eq}p^{Al}_{eq}\
&=& 0,
\\ \label{B-8}
(\lambda^A_{,k}u^A_l+\lambda^A u^A_{l,k})^{eq}\frac{1}{c^2}q^{Ak}_{eq}u^{Al}_{eq}\
&=& 0,
\\ \label{B-9}
-(\lambda^A_{,k}u^A_l+\lambda^A u^A_{l,k})^{eq}p^Ah^{Akl}_{eq}\
&=& 0,
\\ \label{B-10}
(\lambda^A_{,k}u^A_l+\lambda^A u^A_{l,k})^{eq}\pi^{Akl}_{eq}\
&=& 0.
\eey
Summing up \R{B-6} to \R{B-10} yields
\bee{I9}
(\lambda^A u^A_{l}){^{eq}_{,k}}T_{eq}^{Akl}\ =\ 0.
\ee
Consequently, \R{V9} is satisfied because each of the three terms vanishes for its own, thus being compatible with \R{V12}.
If an $^A$-component of a mixture is in equilibrium, two types of equilibria can occur: one in an
arbitrary material showing the usual equilibrium conditions and another one which has beyond the
the usual equilibrium conditions vanishing temperature gradient and vanishing 4-velocity gradient
according to \R{V13}.
\vspace{.3cm}\newline
Evident is that an 1-component system which does not interact with other components is as a
special case included in the theory of an $^A$-component in the mixture. This case is discussed in the next section.

\section{Special Case: Thermodynamics of 1-Components\label{FC}}
\subsection{Entropy flux, -supply and -density}

An 1-component system\footnote{that is not a mixture which is a
multi-component system by definition}
can be described by setting equal all component indices of a multi-component system
\bee{O3x} 
A, B, C,...,Z\quad\longrightarrow\quad 0,
\ee
and for shortness, we omit this common index 0. Then the basic fields of an 1-component
system are according to \R{K3}
\bee{O4x}
\mbox{rest mass density and 4-velocity:}\hspace{.5cm}\{\rh,u_k\}.
\ee
The equations \R{K5} of Setting I change into identities. According to \R{K6b}$_1$, \R{K8c}$_1$,
\R{K9a}$_2$ and \R{O2}, we have
\bee{O5x}
f\ =\ 1,\quad J_k\ =\ 0,\quad w\ =\ 0,\quad T^{kl}{_{,k}}\ =\ k^l.
\ee
The accessory variables become acording to \R{Z3}, \R{Z7} and \R{Z10}
\bee{O6x}
\lambda\ =\ \frac{\nu}{\Theta},\qquad \kappa\ =\ \frac{\mu}{\Theta},\qquad\lambda^k\ =\ 0.
\ee
The entropy density \R{Z14p} and the state space \R{Y2} are as in equilibrium of the
$^A$-component \R{W1}
\bee{O7x}
s\ =\ \frac{1}{\Theta}\Big(\mu\rh +\nu(e + p) \Big),\qquad{\sf z}\ =\ (\rh, e).
\ee
The entropy flux \R{Z15p}, the entropy supply \R{Z16p} and the entropy production \R{T7gy}
are\footnote{There are no chemical reactions in 1-component systems.}
\bee{O8x}
s^k\ =\ \frac{\nu}{\Theta} q^k,\quad \varphi\ =\ \frac{1}{\Theta}\Big(\mu\ {^{(ex)}}\Gamma + \nu u_lk^l\Big),\quad 
\sigma\ =\ \Big(\frac{\nu}{\Theta} u_l\Big)_{,k}\Big(T^{kl}
-\frac{1}{c^2}eu^ku^l+ph^{kl}\Big).
\ee
According to sect.\ref{COMI}, the (3+1)-components of the mixture change into the corresponding
quantities of the 1-component system. The necessary equilibrium conditions of an 1-component
system are equal to those of an $^A$-component in the mixture\footnote{in equilibrium: "all cats are grey"}.

\subsection{Equilibrium and reversibility}

Considering an 1-component system, \R{B1a} results in
\bee{O8x1}
(\lambda u_l)_{,k}\Big(T^{kl}-\frac{1}{c^2}eu^ku^l+ph^{kl}\Big)\ =\ 0.
\ee
In contrast to \R{B1a}, \R{O8x1}
is sufficient and necessary for vanishing entropy production in 1-component systems according to \R{T7gy} and \R{O8x}$_3$. But concerning equilibrium, \R{O8x1} is only
necessary but not sufficient for it, because further equilibrium conditions according to
\R{O8x}$_{1,2}$
\bee{O8x2}
q^k_{eq}\ =\ 0,\qquad u_l^{eq}k^l_{eq}\ =\ 0,\qquad ^{(ex)}\Gamma^{eq}\ =\ 0
\ee
may not be satisfied. Vanishing entropy production out of equilibrium belongs to reversible
processes and vice versa.  Consequently, reversible processes in 1-component systems satisfy
\bee{O8x3}
\sigma^{rev}\ =\ 
(\lambda u_l)_{,k}^{rev}\Big(T^{kl}-\frac{1}{c^2}eu^ku^l+ph^{kl}\Big)^{rev}\ =\ 0
\ee
which has the same structure as \R{V9} in case of equilibrium. Thus, all results of
sect.\ref{G4T} change into those
of an 1-component system, if the component index $^A$ is omitted, {\em eq} is replaced by {\em rev}, equilibrium is not presupposed and the generated expressions belong to reversible processes and vice versa. Consequently, the derivative of the 4-temperature and
the Killing relation of the 4-temperature 
\bee{O23x}
(\lambda u_l)_{,k}^{rev}\ =\ 0,\qquad\mbox{or}\quad T^{kl} = T^{lk}\!:\ 
\Big((\lambda u_l)_{,k}+(\lambda u_k)_{,l}\Big)^{rev}\ =\ 0
\ee 
are rather conditions for reversible processes in 1-component systems because the entropy
production is enforced to be zero without existing equilibrium. Summarized: \R{O8x1} is necessary for equilibrium and additionally sufficient and necessary for reversible processes in 1-component systems.
Independently of the 4-temperature, we obtain the well-known fact \C{STE} that
all processes of perfect materials are reversible in 1-component systems according to \R{O8x}$_3$
\bee{O8xc}
T^{kl}_{per}\ :=\ \frac{1}{c^2}eu^ku^l-ph^{kl}\ \longrightarrow\ \sigma_{per}\ =\ 0.
\ee

\section{Thermodynamics of a Mixture\label{TM}}

According to the mixture axiom in sect.\ref{MIX}, the balance equations of a mixture look like those of an 1-component system. But a mixture as a whole behaves differently from
the interacting $^A$-component in the mixture and also differently from
an 1-component system which both were discussed in sect.\ref{IC} and sect.\ref{FC}.
Because the interaction between the components is still existing in the mixture, the
diffusion fluxes and also the multi-heat relaxation do not vanish as in 1-component
systems. Because component indices $^A$ do not appear in the
description of mixtures, they are summed up in contrast to 1-component systems for which they vanish. The Settings I and II enforce the mixture axiom resulting in
\byy{Q-1}
\mbox{mass balance:}&\quad& N^{Ak}{_{,k}}\ =\ \Gamma^A\ \longrightarrow\ 
N^{k}{_{,k}}\ =\ \Gamma,
\\ \label{Q-2}
\mbox{energy balance:}&\quad& u^A_lT^{Akl}{_{,k}}ß =\ \Omega^{A}\ \longrightarrow\ 
u_l{\sf T}^{kl}{_{,k}}\ =\ \Omega,
\\ \label{Q-3}
\mbox{momentum balance:}&\quad& h^{Am}_lT^{Akl}{_{,k}}ß =\ \Omega^{Am}\
\longrightarrow\ h^{m}_l{\sf T}^{kl}{_{,k}}\ =\ \Omega^m,
\eey
According to \R{K14g}, \R{K14h} and \R{O2}, we obtain
\bee{Q-4a}
u_lk^l\ =\ \sum_A\Big\{\Big(h^{Am}_lu_m +
f^Au^A_l\Big)k^{Al}\Big\},
\quad
h^p_lk^l\ =\ \sum_A\Big\{\Big(h^{Am}_lh^p_m +
g^{Ap}u^A_l\Big)k^{Al}\Big\}.\hspace{.8cm}
\ee
Consequently, the mixture axiom --represented by \R{Q-1} to \R{Q-3}-- does not demand
in contrast to the non-relativistic case the additivity of the balances of the
$^A$-components. 
\vspace{.3cm}\newline
Setting III to Setting VIII are related to the $^A$-component itself, whereas Setting I and
II describe additivity properties in the set of the $^A$-components. Both kinds of setting do not contain the entropy of the mixture. Obviously, we need an additional setting concerning the entropy of the mixture which will be formulated in the next section.

\subsection{Additivity of 4-entropies}
\subsubsection{Entropy density and -flux}

Like the additivity of the mass flux densities \R{K5}$_1$ and the energy-momentum tensors
\R{K11c}$_2$  of the $^A$-components, we demand that also of the 4-entropies are additive
\bey\nonumber
\mbox{\sf Setting X:}\hspace{6cm}
\\ \label{Q1}
{\sf S}^k\ \st{\td}{=}\ \sum_ A S^{Ak}.   \hspace{3cm}
\eey
Consequently, we obtain from \R{E2} 
\bee{E3}
{\sf S}^k\ =\ \sum_A\Big\{
\K{A}\Big(\nu^AQ^{Ak}+u_m\tau^{Akm}\Big)
+\frac{\mu^A}{\Theta^A}\Big((1+w^A)N^{Ak}-J^{Ak}\Big)
+\frac{\nu^A}{\Theta^A}p^Au^{Ak}+G^{Ak}\Big\}.
\vspace{.3cm}\ee
According to \R{T2z}, we obtain the entropy density and the entropy flux density
of the mixture by use of \R{K9}, \R{K6b}$_1$ and \R{L1}
\bey\nonumber
{\sf S}^ku_k = c^2{\sf s} &=&
\sum_A\Big\{
\K{A}\Big(\nu^AQ^{Ak}+u_p\tau^{Akp}\Big)u_k+
\\ \label{E4}
&&\hspace{.7cm}+\frac{\mu^A}{\Theta^A}\Big(1+w^A\Big)\rh^Ac^2f^A
+\frac{\nu^A}{\Theta^A}p^Ac^2f^A+G^{Ak}u_k\Big\},
\\ \nonumber
{\sf S}^kh_k^m = {\sf s}^m &=&
\sum_A\Big\{
\K{A}\Big(\nu^AQ^{Ak}+u_p\tau^{Akp}\Big)h_k^m+
\\ \label{E5}
&&\hspace{.7cm}+\frac{\mu^A}{\Theta^A}w^AJ^{Am}
+\frac{\nu^A}{\Theta^A}p^Ac^2g^{Am}+G^{Ak}h_k^m\Big\}.\hspace{2cm}
\vspace{.3cm}\eey
Taking \R{K11d} into consideration, we introduce by comparing with \R{E4} and \R{E5} the
\bey\nonumber
\mbox{\sf Setting XI:}\hspace{5cm}
\\ \label{E6}
\nu^A\ \st{\td}{=}\ f^A .   \hspace{3cm}
\eey
With this setting, the expressions of the entropy density and the entropy flux of the
mixture correspond to those which are generated by the additivity of the
energy-momentum tensors: \R{K15b} to \R{K15j}. 
\vspace{.3cm}\newline
Finally, we obtain the entropy and entropy flux density of the mixture
\bey\nonumber
{\sf s} &=&
\sum_A\Big\{\frac{1}{c^2}
\K{A}\Big(f^AQ^{Ak}+u_p\tau^{Akp}\Big)u_k+
\\ \label{E7}
&&\hspace{.7cm}+\frac{\mu^A}{\Theta^A}\Big(1+w^A\Big)\rh^Af^A
+\frac{f^A}{\Theta^A}p^Af^A+\frac{1}{c^2}G^{Ak}u_k\Big\},
\\ \nonumber
{\sf s}^m &=&
\sum_A\Big\{
\K{A}\Big(f^AQ^{Ak}+u_p\tau^{Akp}\Big)h_k^m+
\\ \label{E8}
&&\hspace{.7cm}+\frac{\mu^A}{\Theta^A}w^AJ^{Am}
+\frac{f^A}{\Theta^A}p^Ac^2g^{Am}+G^{Ak}h_k^m\Big\}.\hspace{2cm}
\eey
These expressions of the entropy and entropy flux densites of the mixture are direct results of Setting X \R{Q1}. They will be discussed in sect.\ref{EFT}.

\subsubsection{Entropy production density and -supply}

From \R{Q1} and \R{T5} follows the entropy balance equations of the mixture
\bee{Q4}
{\sf S}^k{_{,k}}\ =\ \sum_ A S^{Ak}{_{,k}}\ =\ \sum_A\Big(\sigma^A+\varphi^A\Big)\ =\ 
^\dm\sigma +^\dm\!\varphi,
\ee
satisfying  the mixture axiom. Accepting the additivity of the entropy supplies of the
$^A$-components\footnote{Supplies are caused by external influences, productions by
internal ones. That the reason why they do not mix up.}
\bey\nonumber
\mbox{\sf Setting XII:}\hspace{6cm}
\\ \label{Q5}
^\dm\!\varphi\ \st{\td}{=}\ \sum_A\varphi^A,\hspace{3.2cm}
\eey
we obtain the additivity of the entropy productions  of the $^A$-components
\bee{Q6}
^\dm\sigma\ =\ \sum_A\sigma^A.\hspace{1.6cm}
\vspace{.3cm}\ee
We obtain the entropy supply of the mixture from \R{Z16p}, \R{Q5} and \R{E6}
\bee{Q7}
^\dm\!\varphi\ =\ \sum_A\Big\{\frac{1}{\Theta^A}\Big(\mu^A{^{(ex)}}\Gamma^A+
f^A u_l^Ak^{Al} + u_ph^{Ap}_mk^{Am}\Big)+\sum_B H^{ABk}{_{,k}}\Big(\frac{1}{\Theta^A}-\frac{1}{\Theta^B}\Big)\Big\}.
\ee
The entropy production of the mixture follows from \R{T9zp}, \R{Q6} and \R{E6}
\bey\nonumber
^\dm\!\sigma\ =\ \sum_A\Big\{
\frac{\mu^A}{\Theta^A}\Big({^{(in)}}\Gamma^A
-(J^{Am}h^{Ak}_m)_{,k}\Big)
-\Big(\frac{\mu^A}{\Theta^A}\Big)_{,k}J^{Am}h^{Ak}_m +
\\ \nonumber
+\ \Big(\frac{f^A}{\Theta^A}\Big)_{,k}q^{Ak}
+\ \frac{f^A}{\Theta^A}u^A{_{l,k}}\Big(\pi^{Akl}
+u^{Ak}p^{Al}\Big)
+\sum_B H^{ABk}\Big(\frac{1}{\Theta^A}-\frac{1}{\Theta^B}\Big)_{,k}+
\\ \label{Q8}
+\ \Big(\frac{1}{\Theta^A}u^mh^A_{ml}\Big)_{,k}
\Big(\frac{1}{c^2}e^Au^{Ak}u^{Al}
+\frac{1}{c^2}q^{Ak}u^{Al}+t^{Akl}\Big)\Big\}.
\eey

\subsection{Entropy, entropy flux densities and partial temperatures\label{EFT}}

We now consider the two expressions
\bee{Q2k}
\frac{1}{c^2}u_k\sum_A\frac{1}{\Theta^A}\Big(Q^{Ak}f^A
+ u_p \tau^{Akp}\Big)\quad\mbox{and}\quad
h^m_k\sum_A\frac{1}{\Theta^A}\Big(Q^{Ak}f^A+ u_p \tau^{Akp}\Big)
\ee
which appear in the entropy density \R{E4} and in the entropy flux density \R{E5}. A comparison with \R{K15b} and \R{K15d} 
\byy{Q2l}
\frac{1}{c^2}u_k\sum_A\Big(Q^{Ak}f^A+ u_p \tau^{Akp}\Big) &=& {\sf e},
\\ \label{Q2m}
h^m_k\sum_A\Big(Q^{Ak}f^A+ u_p \tau^{Akp}\Big)&=& {\sf q}^m
\eey
results in the fact, that the partial temperatures $\Theta^A$ in \R{Q2k} prevent from introducing the mixture quantities energy density \R{Q2l} and the energy flux density
\R{Q2m} into the entropy density \R{E7} and the entropy flux density \R{E8} of the
mixture. 
\vspace{.3cm}\newline
Beyond that, the temperature of the mixture $^\dm\Theta$ cannot be defined properly: even the mean value theorem does not help
\byy{Q2n} 
\frac{1}{c^2}u_k\sum_A\frac{1}{\Theta^A}\Big(Q^{Ak}f^A+ u_p \tau^{Akp}\Big) &\st{?}{=}&
\frac{1}{c^2}\frac{1}{^\dm\,\Theta}u_k\sum_A\Big(Q^{Ak}f^A+ u_p \tau^{Akp}\Big) = \frac{1}{^\dm\,\Theta}{\sf e},
\\ \label{Q2o}
h^m_k\sum_A\frac{1}{\Theta^A}\Big(Q^{Ak}f^A+ u_p \tau^{Akp}\Big)&\st{?}{=}& 
\frac{1}{^\dm\,\Theta}(m)h^m_k\sum_A\Big(Q^{Ak}f^A+ u_p \tau^{Akp}\Big) = \frac{1}{^\dm\,\Theta}(m){\sf q}^m.
\hspace{.9cm}
\eey
In the first case \R{Q2n}, it is not for sure that the presupposition
\bee{Q2p}
u_k\Big(Q^{Ak}f^A+ u_p \tau^{Akp}\Big)\quad\mbox{definite for all $A$} 
\ee
is satisfied which is sufficient
for the validity of the mean value theorem. The second case \R{Q2o} results in mean
values which may depend on the tensor component index, quantities which are useless for defining a common temperature.
\vspace{.3cm}\newline
There exists another argumentation that a temperature of a mixture cannot be defined properly: the partial temperatures are internal contact variables \C{MU14} measured by
thermometers which are selective for the temperature $\Theta^A$ of the corresponding
$^A$-component. Evident is, that the temperature of the mixture is a certain mean value
of the partial temperatures of the components of the mixture \C{MUE68,DUMUE,BOGA}
\bee{Q2q}
^\dm\,\Theta\ =\ \sum_B\alpha_B\Theta^B,\quad\alpha_B\geq 0,\quad\sum_B\alpha^B\ =\ 1.
\ee
But this mean value depends on the selectivity of the used thermometer, and consequently
the measured temperature is not unequivocal. Different definitions of the mixture
temperature can be found in literature \C{CAJO}.
But a unique mixture temperature --independent of thermometer selectivities or arbitrary definitions--
is given in the case of {\em multi-heat relaxation equilibrium} \R{W1a}. This case is silently presupposed, if only one temperature is used in multi-component non-equilibrium systems. Only this sure case is considered in the sequel.

\subsection{Multi-heat relaxation equilibrium\label{MHC}}
\subsubsection{Entropy and entropy flux densities}

Presupposing  multi-heat relaxation equilibrium \R{W1a}, 
\bee{Q3}
G^{ABk}_{eq}\ \doteq\ 0
\ee
follows according to \R{A10}, and the expressions \R{Q2k} transfer according to \R{Q2l} and \R{Q2m} to
\byy{Q3a}
\frac{1}{c^2}u_k\sum_A\frac{1}{\Theta^A}\Big(Q^{Ak}f^A+ u_p \tau^{Akp}\Big)
&=& \frac{1}{^\dm\,\Theta}{\sf e},
\\ \label{Q3b}
h^m_k\sum_A\frac{1}{\Theta^A}\Big(Q^{Ak}f^A+ u_p \tau^{Akp}\Big)&=& 
\frac{1}{^\dm\,\Theta}{\sf q}^m.
\eey
Entropy density \R{E7} and entropy flux density \R{E8} become
\byy{Q3c}
{\sf s} &=&\frac{1}{^\dm\,\Theta}{\sf e}+
\frac{1}{^\dm\,\Theta}\sum_A\Big\{\mu^A\rh^A\Big(1+w^A\Big)
+p^Af^A\Big\}f^A,
\\ \label{Q3d}{\bullet}\hspace{3cm}
{\sf s}^m &=&\frac{1}{^\dm\,\Theta}{\sf q}^m
+\frac{1}{^\dm\,\Theta}\sum_A\Big\{\mu^Aw^AJ^{Am}
+f^Ap^Ac^2g^{Am}\Big\}.\hspace{3cm}
\eey
The first term in the sum of \R{Q3c} can be exploited by use of the mean value theorem
because in contrast to \R{Q2p}, the presupposition for its validity is here satisfied according to \R{K6b}$_2$
\bee{Q3e}
\sum_A\mu^A\rh^A f^A\ =\ ^\dm\mu\sum_Af^A\rh^A\ =\ 
^\dm\mu\rh.
\ee
Consequently, the chemical potential of the mixture is
\bee{Q3f}
{^\dm}\, \mu\ :=\ \sum_A\mu^A f^A\frac{\rh^A}{\rh},
\ee
and the entropy density of the mixture \R{Q3c} yields
\bee{Q3g}
{\sf s}\ =\ \frac{1}{^\dm\,\Theta}{\sf e}+\frac{1}{^\dm\,\Theta}{^\dm}\mu\rh
+\frac{1}{^\dm\,\Theta}\sum_A\Big\{\mu^A\rh^Aw^Af^A
+p^A(f^A)^2\Big\}
\ee
Introducing an effective "pressure"
\bee{Q3h}
{\sf p^{eff}}\ :=\ \sum_A\Big\{\mu^A\rh^Aw^Af^A+p^A(f^A)^2\Big\}
\ee
which is different from ${\sf p}$ in \R{K15j1}\footnote{${\sf p}$
does not depend on $\rh^A$ and $\mu^A$, whereas $\sf p^{eff}$ is independent of
$e^A,\ q^{Ak},\ p^{Ak}$ .}.
The entropy density of the mixture in
multi-heat relaxation equilibrium has the usual shape \R{O7x} as for an 1-component
system\footnote{$\nu =f=1$}
\bee{Q3i}\bullet\hspace{5cm}
{\sf s}\ =\ \frac{1}{^\dm\,\Theta}\Big( {^\dm}\mu\rh + {\sf e} + {\sf p^{eff}}\Big).
\hspace{5cm}
\ee
We now consider the equilibrium values of ${\sf p}$ and ${\sf p^{eff}}$. Starting
with \R{K15j1} and taking \R{Y16d} and \R{P23c}$_{2,3}$ into account, we obtain
\bee{Q3i0}
{\sf p_{eq}}\ =\ -\frac{1}{3}\sum_At^{Akl}h^A_{kl}\ =\ \sum_ Ap^A\ =\
{\sf p_{eq}^{eff}},
\ee
that means, in equilibrium appears the pressure of the mixture in its entropy density,
whereas in non-equilibrium this pressure is replaced by an effective quantity which has
the physical dimension of a pressure, but which is different from it.

\subsubsection{Entropy production, -density and -supply}

In multi-heat relaxation equilibrium, entropy supply \R{Q7} and entropy production
density \R{Q8} of the mixture become
\byy{Q9}\bullet\hspace{2cm}
^\dm\!\varphi &=& \frac{1}{^\dm\,\Theta}\sum_A\Big(\mu^A{^{(ex)}}\Gamma^A+
f^A u_l^Ak^{Al} + u_ph^{Ap}_mk^{Am}\Big),
\\ \nonumber\bullet\hspace{2cm}
^\dm\!\sigma &=& \sum_A\Big\{
\frac{\mu^A}{^\dm\,\Theta}\Big({^{(in)}}\Gamma^A
-(J^{Am}h^{Ak}_m)_{,k}\Big)
-\Big(\frac{\mu^A}{^\dm\,\Theta}\Big)_{,k}J^{Am}h^{Ak}_m +
\\ \nonumber
&&\hspace{.7cm}+\ \Big(\frac{f^A}{^\dm\,\Theta}\Big)_{,k}q^{Ak}
+\ \frac{f^A}{^\dm\,\Theta}u^A{_{l,k}}\Big(\pi^{Akl}
+u^{Ak}p^{Al}\Big)+
\\ \label{Q10}
&&\hspace{.7cm}+\ \Big(\frac{1}{^\dm\,\Theta}u^mh^A_{ml}\Big)_{,k}
\Big(\frac{1}{c^2}e^Au^{Ak}u^{Al}
+\frac{1}{c^2}q^{Ak}u^{Al}+t^{Akl}\Big)\Big\}.\hspace{2cm}
\eey
The meaning of each individual term of \R{Q10} was already discussed with regard to
the $^A$-component according to \R{T9zp}.
\vspace{.3cm}\newline
Entropy density \R{Q3i}, entropy flux density \R{Q3d}, entropy supply \R{Q9},
entropy production density \R{Q10} and chemical potential \R{Q3f} of the mixture
are represented by sums of quantities of the $^A$-components. As expected, the
(3+1)-components of the energy-momentum tensors cannot represent the mentioned
thermodynamical quantities because diffusion fluxes, chemical potentials and temperature
are not included in the energy-momentum tensor. From them, only the energy density
${\sf e}$ and the energy flux density ${\sf q}^m$ of the mixture appear in entropy and
entropy flux densities. All other thermodynamical quantities, such as the chemical
potential $^\dm\,\mu$ \R{Q3f} and the effective pressure ${\sf p^{eff}}$ \R{Q3h}
 are composed of
$^A$-component quantities which do not belong to the energy-momentum tensor.
A temperature $^\dm\,\Theta$ of the mixture exists in multi-heat relaxation equilibrium
by definition, but it is not composed by the partial temperatures $\Theta^A$ of the
$^A$-components but rather by the component sensitivity of a thermometer.

\subsection{Equilibrium}

Evident is that the equilibrium conditions of a mixture follow from those of the
$^A$-com\-po\-nents which we considered in sect.\ref{EQ}. Consequently, to demand
additional equilibrium conditions for mixtures is not necessary. Starting with \R{Q3},
we obtain with \R{P23}$_2$, \R{P23d}$_3$, \R{P23c}$_2$ and
\R{W1}\footnote{$\nu^A_{eq}=f^A_{eq}=1$}
\bee{O18}
{\sf s}^m_{eq}\ =\ 0,\qquad {\sf s}_{eq}\ =\ \sum_A s^A_{eq}\ =\ \sum_A
\lambda^A_{eq}\Big(\mu^A_{eq}\rh^A _{eq}+e^A_{eq} + p^A_{eq}\Big)
\ee
According to \R{Q5}, \R{Q6} and \R{P23}$_{1,3}$, we obtain
\bee{O19}
^\dm\varphi_{eq}\ =\ 0,\qquad ^\dm\sigma_{eq}\ =\ 0.
\ee
From \R{K15e} follows with \R{P23d}$_1$ and \R{O18}$_1$
\bee{O20}
{\sf q}^m_{eq}\ =\ 0.
\ee
From \R{K15c}, \R{K15g} and \R{K15j} follows with \R{P23d}$_1$ and \R{Y16d}
\bee{O21}
{\sf e}_{eq}\ =\ \sum_A e^A_{eq}\qquad 
{\sf p}^m_{eq}\ =\ \sum_A p^{Am}_{eq},\qquad {\sf t}^{jm}_{eq}\ =\ \sum_At^{Ajm}_{eq}.
\ee
According to \R{K7a}$_3$, \R{K3a} and \R{Y15a}, we obtain
\bee{O22}
^{(ex)}\Gamma_{eq}\ =\ 0,\quad ^{(in)}\Gamma_{eq}\ =\ 0,\quad \Gamma_{eq}\ =\ 0.
\ee
From \R{K4} follows in equilibrium analogously to \R{Y16}
\bee{O23}
u^k_{eq}{_{\ ,k}}\ =\ 0.
\ee
The vanishing entropy supply of the mixture \R{O19}$_1$ is according to \R{Q9}
satisfied by the equilibrium conditions \R{Y15a}$_1$, \R{W2} and \R{Y16d}.
The vanishing entropy production of the mixture \R{O19}$_2$ is according to \R{Q10}
satisfied by the equilibrium conditions \R{Y15a}$_2$, \R{P23a}$_1$, \R{Y16a}, \R{Y16b}, \R{P24} and \R{Y16d}.

\section{Verification: The non-relativistic case}

The non-relativistic case\footnote{Do not take this case for the post-Newtonian limit. Here
we are not interested in relativistic corrections of the non-relativistic equations as done in \C{BOCH,CHA,SCH}.
Here we need the non-relativistic equations for a verification of the relativistic ones.} is defined by the formal setting\footnote{$\ap$ characterizes
the approximation step which results in the non-relativistic Newtonian limit.}
\bee{A10x}
\frac{\mvec{v}}{c}\ \ap\ \mvec{0},\quad \mbox{or more formally:}\ c\ \longrightarrow\ 
\infty
\ee
which generates the non-relativistic Newtonian shape of the thermodynamical expressions of sect.\ref{MHC}.
\vspace{.3cm}\newline
The approximation \R{A10x} represents the zeroth step of developing the relativistic
squareroot
\bee{A10x1}
\sqrt{1-(v/c)^2}\ \ap\ 1.
\ee
The energy-momentum tensor \R{J1} and the 4-entropy \R{T2} are chosen in such a way
that their (3+1)-components are all of the same order of approximation, namely of zeroth
order. Consequently, a "book keeping" \C{SCH,WIL,SCHBOCHMU}
fixing different conditions of
approximation orders for these (3+1)-components is not required in the case of zeroth
approximation generating the Newtonian limit of continuum thermodynamics.

\subsection{Velocity and mass balance}

The norm of the 4-velocity (k=1,...4; $\alpha$=1,...3)
\bee{Q1x}
u^ku_k\ =\ u^\alpha u_\alpha + u^4u_4\ =\
 u_mg^{mk}u_k\ =\  u_\alpha g^{\alpha\beta}u_\beta
+ u_\alpha g^{\alpha 4}u_4 + u_4g^{4\beta}u_\beta +u_4g^{44}u_4. 
\ee
can be decomposed into its space-like, time-like and mixed parts ($\alpha =1,2,3)$. We now introduce two frames having a special Galilean metric\footnote{$\bg$ marks the introduction of Galilean co-ordinates (which may have different signature.}
\bee{Q2y}
g^{\alpha\beta}\ \bg\ 
\mp \delta^{\alpha\beta},\quad g^{44}\ \bg\ \pm 1,\quad
g^{\alpha 4}\ = \ g^{4\alpha}\ \bg\ 0,\ \longrightarrow\ \mbox{signature}(g^{mk}) = \mp 2. 
\ee
The corresponding parts of the 4-velocity are chosen as\footnote{a component index
$^A$ is suppressed}
\byy{Q2xy}
u^\alpha &\bg& \frac{v^{\alpha}}{\sqrt{1-(v/c)^2}},\qquad
u_\alpha\ \bg\ \mp\frac{v_{\alpha}}{\sqrt{1-(v/c)^2}},
\\ \label{Q3y}
u^4 &\bg& \frac{c}{\sqrt{1-(v/c)^2}},\qquad
u_4\ \bg\ \pm\frac{c}{\sqrt{1-(v/c)^2}}.
\eey
Consequently, \R{Q1x} results in
\bee{Q4y}
u^ku_k\ =\ u^\alpha u_\alpha + u^4u_4\ \bg\
\mp\frac{\mvec{v}\cdot\mvec{v}}{1-v^2/c^2}
\pm\frac{c^2}{1-v^2/c^2}\ =\ \pm c^2.\quad\surd
\vspace{.3cm}\ee
By taking
\bee{U7c}
\p\Big(1-(v/c)^2\Big)^{-1/2}\ =\ 
-\frac{1}{2}\Big(1-(v/c)^2\Big)^{-3/2}\p\Big(\frac{v}{c}\Big)^2\ \ap\ 0
\ee
into account, the relativistic mass balance changes into the usual non-relativistic one
\bey\nonumber
N^k{_{,k}} &=& (\rh u^k )_{,k}\ =\ (\rh u^\alpha )_{,\alpha} + (\rh u^4 )_{,4}\ \bg
\\ \nonumber
&\bg& \nabla\cdot\Big(\rh\frac{\mvec{v}}{\sqrt{1-(v/c)^2}}\Big)
+\frac{1}{c}\p\Big(\rh\frac{c}{\sqrt{1-(v/c)^2}}\Big)\ =\ \Gamma\ap
\hspace{1cm}\mbox{}
\\ \label{U-2}
\bullet\hspace{4cm}
&\ap&  \nabla\cdot(\rh\mvec{v})
+\p(\rh)\ =\ \Gamma.
\eey

\subsection{The momentum balance}

Because the energy-momentum tensors of the mixture \R{K11c}, of one component in
the mixture \R{K11a} and of a 1-component system have the same shape, we can start
with \R{K11a} and \R{K11b} without using the component index $^A$
\bee{G1}
T^{kl}{_{,k}}\ =\ \frac{1}{c^2}\Big(q^{k}+eu^{k}\Big)_{,k}u^l
+\frac{1}{c^2}\Big(q^{k}+eu^{k}\Big)u^l{_{,k}}
+ \Big(t^{kl}+u^kp^l\Big)_{,k},
\ee
resulting in
\bee{G2}
h^m_lT^{kl}{_{,k}}\ =\ \frac{1}{c^2}\Big(q^{k}+eu^{k}\Big)u^l{_{,k}}h^m_l
+\Big(t^{kl}+u^kp^l\Big)_{,k}\Big(\delta^m_l\mp\frac{1}{c^2}u^mu_l\Big).
\ee
Inserting
\bee{G3}
u^l{_{,k}}h^m_l\ =\ u^m{_{,k}},
\ee
the RHS of the momentum balance equation \R{G2} becomes
\bee{G4}
h^m_lT^{kl}{_{,k}}\ =\ \frac{1}{c^2}\Big(q^{k}+eu^{k}\Big)u^m{_{,k}}
+\Big(t^{km}+u^kp^m\Big)_{,k}
\mp\frac{1}{c^2}\Big(t^{kl}+u^kp^l\Big)_{,k}u_lu^m.
\vspace{.3cm}\ee
In Galilean co-ordinates, we have for the first term of \R{G4}
\bee{G5}
\frac{1}{c^2}\Big(q^{k}+eu^{k}\Big)u^m{_{,k}}
\ \bg\
\frac{1}{c^2}\Big(q^{\alpha}+e\frac{v^{\alpha}}
{\sqrt{1-(v/c)^2}}\Big)u^m{_{,\alpha}}
+\frac{1}{c^2}\Big(q^{4}+e\frac{c}{\sqrt{1-(v/c)^2}}\Big)u^m{_{,4}}.
\ee
According to \R{J3a}$_1$, we have with \R{Q2xy}$_2$ and \R{Q3y}$_2$
\bee{U4}
q^ku_k\ =\ 
0\ \bg\ \mp\mvec{q}\cdot\frac{\mvec{v}}{\sqrt{1-(v/c)^2}}
\pm q^4\frac{c}{\sqrt{1-(v/c)^2}}
\quad\longrightarrow\quad q^4\ =\ \mvec{q}\cdot\frac{\mvec{v}}{c}\ \ap\ 0.
\ee
According to \R{U7c}, the following non-relativistic limits are valid
\bee{G6}
u^\beta{_{,\alpha}}\ \ap\ \nabla\mvec{v},\quad
u^4{_{,\alpha}}\ \ap\ \nabla c\ =\ \mvec{0},\quad
u^\alpha{_{,4}}\ \ap\ \frac{1}{c}\p \mvec{v}\ =\ \mvec{0},\quad
u^4{_{,4}}\ \ap\ \frac{1}{c}\p c\ =\ 0.
\ee
Consequently, we obtain for the first term of \R{G4} according to \R{G5} to \R{G6}
\byy{G6a}
&&\frac{1}{c^2}\Big(q^{\alpha}+eu^{\alpha}\Big)u^\beta{_{,\alpha}}\ \ap\ 0,\quad
\frac{1}{c^2}\Big(q^{4}+eu^{4}\Big)u^\beta{_{,4}}\ \ap\ 0,
\\ \label{G6b}
&&\frac{1}{c^2}\Big(q^{\alpha}+eu^{\alpha}\Big)u^4{_{,\alpha}}\ \ap\ 0,\quad
\frac{1}{c^2}\Big(q^{4}+eu^{4}\Big)u^4{_{,4}}\ \ap\ 0,
\\ \label{G7}
\longrightarrow\quad&&\frac{1}{c^2}\Big(q^{k}+eu^{k}\Big)u^m{_{,k}}\ \ap\ 0.
\vspace{.3cm}\eey
The second term of \R{G4} is in Galilean co-ordinates  
\bee{G8}
\Big(t^{km}+u^kp^m\Big)_{,k}\ \bg\
\Big(t^{\alpha m}+\frac{v^{\alpha}}{\sqrt{1-(v/c)^2}} p^m\Big)_{,\alpha}
+\Big(t^{4m}+\frac{c}{\sqrt{1-(v/c)^2}}p^m\Big)_{,4}.
\ee
Starting with \R{J3a}$_3$ and \R{J3a}$_4$, we obtain by taking \R{Q2xy}$_2$ and \R{Q3y}$_2$ into account
\byy{U6}
u_kt^{kl}\ =\
0 &\bg& \mp\frac{v_{\alpha}}{\sqrt{1-(v/c)^2}}t^{\alpha l} \pm\frac{c}{\sqrt{1-(v/c)^2}}t^{4l}
\ \longrightarrow\ t^{4l} = \frac{v_\alpha}{c} t^{\alpha l}\ \ap\ 0,
\\ \label{U7}
t^{kl}u_l\ =\
0 &\bg& \mp\frac{v_{\alpha}}{\sqrt{1-(v/c)^2}}t^{k\alpha} \pm\frac{c}{\sqrt{1-(v/c)^2}}t^{k4}
\ \longrightarrow\ t^{k4} = \frac{v_\beta}{c} t^{k\beta}\ \ap\ 0.
\eey
Especially, setting $l=\beta,4$ and $k=\alpha$ in \R{U6} and \R{U7}, we obtain
\bee{U7a}
t^{44}\ =\ \frac{\mvec{v}}{c}\cdot{\bf t}\cdot\frac{\mvec{v}}{c}\ \ap\ 0.
\ee
According to \R{J3a}$_2$, we have
\bee{U4z}
p^ku_k\ =\ 
0\ \bg\ \mp\mvec{p}\cdot\frac{\mvec{v}}{\sqrt{1-(v/c)^2}} \pm p^4\frac{c}{\sqrt{1-(v/c)^2}}
\quad\longrightarrow\quad p^4\ =\ \mvec{p}\cdot\frac{\mvec{v}}{c}\ \ap\ 0.
\ee
Taking \R{U6} to \R{U4z} into account, the non-relativistic limit of the second term of
\R{G4} becomes according to \R{G8}
\byy{G9}
&&\Big(t^{\alpha\beta}+u^\alpha p^\beta\Big)_{,\alpha}\ \ap\
\nabla\cdot\Big({\bf t}+\mvec{v}\mvec{p}\Big),\quad
\Big(t^{4\beta}+u^4p^\beta\Big)_{,4}\ \ap\ \p\mvec{p},
\\ \label{G9a}
&&\Big(t^{\alpha4}+u^\alpha p^4\Big)_{,\alpha}\ \ap\ 0,\hspace{2.7cm}
\Big(t^{44}+u^4p^4\Big)_{,4}\ \ap\ 0,
\\ \label{G9b}
\longrightarrow\quad&&\Big(t^{km}+u^kp^m\Big)_{,k}\ \ap\ \p\mvec{p}+
\nabla\cdot\Big({\bf t}+\mvec{v}\mvec{p}\Big).
\vspace{.3cm}\eey
The third term of \R{G4} is by taking \R{G9a} into account
\bee{G10}
\mp\frac{1}{c^2}\Big(t^{kl}+u^kp^l\Big)_{,k}u_lu^m\ =\
\mp\frac{1}{c^2}\Big[\Big(t^{\alpha\beta}+u^\alpha p^\beta\Big)_{,\alpha}
+\Big(t^{4\beta}+u^4p^\beta\Big)_{,4}\Big]u_\beta u^m.
\ee
Because of
\bee{G11}
\frac{1}{c^2}u_\beta u^m\ \ap\ 0,
\ee
the non-relativistic limit of the third term of \R{G4} vanishes, and we obtain from \R{G7}, \R{G9}
and \R{G9a}
\byy{G13}
h^\alpha_lT^{kl}{_{,k}}&\ap&
\p\mvec{p}+\nabla\cdot\Big(\mvec{v}\mvec{p}+{\bf t}\Big),
\\ \label{G13a}
h^4_lT^{kl}{_{,k}}&\ap&0.
\vspace{.3cm}\eey
We now consider the RHS of the momentum balance equation \R{K14} by taking \R{G11} into account
\byy{G14}
h^m_lT^{kl}{_{,k}}\ =\ h^m_lk^l\
&=& k^m\mp\frac{1}{c^2}u^mu_lk^l,
\\ \label{G15}
 h^\alpha_lk^l&=&k^\alpha\mp\frac{1}{c^2}u^\alpha u_lk^l\ \ap\ k^\alpha,
\\ \label{G16}
h^4_lk^l&=&k^4\mp\frac{1}{c^2}u^4 \Big( u_\alpha k^\alpha + u_4k^4\Big)\
\ap\ k^4-k^4\ =\ 0,
\eey
corresponding to \R{G13a}. Consequently, the non-relativistic momentum balance equation
\bee{G17}{\bullet}\hspace{5cm}
\p\mvec{p}+\nabla\cdot\Big(\mvec{v}\mvec{p}+{\bf t}\Big)\ =\ \mvec{k}
\hspace{5cm}\mbox{}
\ee
is decoupled from the energy balance equation and
--as expected-- is independently valid of the signature of the Galilean co-ordinates. 
The RHS of \R{G17} $\mvec{k}$ is the external force density. The corresponding
4-component $k^4$ is not restricted by the momentum balance equation according to \R{G16}.
\vspace{.3cm}\newline
If the energy momentum-tensor is symmetric, we have according to \R{J3c}
\bee{G18}
T^{kl}\ =\ T^{lk}\ \longrightarrow\ p^m\ =\ \frac{1}{c^2}q^m\ \ap\ 0.
\ee
Inserting \R{G18} into \R{G9b}, the momentum balance equation \R{G17} results in the equilibrium condition
\bee{G19}{\bullet}\hspace{4cm}
T^{kl}\ =\ T^{lk}\ \longrightarrow\ \nabla\cdot{\bf t}\ =\ \mvec{k},\
\quad\ {\bf t}\ =\ {\bf t}^\top.\hspace{3cm}\mbox{}
\ee

\subsection{The energy balance}

Starting with \R{G1}, we obtain the energy balance equation by taking \R{J3a}$_{2,4}$
into account
\bee{U1}
u_lT^{kl}{_{,k}}\ =\ \pm\Big(q^k+eu^k\Big)_{,k}+\Big(t^{kl}+u^kp^l\Big)_{,k}u_l\ 
=\ u_lk^l.
\ee
Decomposing the first term of \R{U1}, we obtain by use of \R{Q2xy}$_{1}$,
\R{Q3}$_{1}$ and \R{U4z}$_{3}$
\bey\nonumber
\pm (q^{k}+eu^{k})_{,k}&=&\pm (q^{4}+eu^{4})_{,4}
\pm(q^{\alpha}+eu^{\alpha})_{,\alpha}\ \bg\
\\ \nonumber
&\bg& \pm\frac{1}{c}\partial_t\Big(\mvec{q}\cdot\frac{\mvec{v}}{c}
+e\frac{c}{\sqrt{1-(v/c)^2}}\Big)
\pm\nabla\cdot\Big(\mvec{q}+e\frac{\mvec{v}}{\sqrt{1-(v/c)^2}}\Big)\ \ap
\\ \label{U5}
&\ap& \pm\Big[\p e+\nabla\cdot\Big(\mvec{q}+e\mvec{v}\Big)\Big].
\vspace{.3cm}\eey
According to \R{G9a}, the second term of \R{U1} results in by taking \R{G9} and \R{G9a} into account
\bey\nonumber 
\Big(t^{kl}+u^kp^l\Big)_{,k}u_l &=&
\Big[\Big(t^{\alpha\beta}+u^\alpha p^\beta\Big)_{,\alpha} 
+\Big(t^{4\beta}+u^4p^\beta\Big)_{,4}\Big]u_\beta\ +
\\ \nonumber
&&+\Big[\Big(t^{\alpha 4}+u^\alpha p^4\Big)_{,\alpha} 
+\Big(t^{44}+u^4p^4\Big)_{,4}\Big]u_4\ \ap
\\ \nonumber
&\ap&\mp\Big[\nabla\cdot\Big({\bf t}+\mvec{v}\mvec{p}\Big)+\p\mvec{p}\Big]\cdot
\mvec{v}\ \pm
\\ \label{U5a}
&&\pm\Big[\nabla\cdot\Big({\bf t}\cdot\mvec{v}+\mvec{v}(\mvec{p}\cdot\mvec{v})\Big)
+\p (\mvec{p}\cdot\mvec{v})\Big].
\eey
Using the momentum balance equation \R{G17}, the second term of \R{U1} results in
\bee{aU5a}
\Big(t^{kl}+u^kp^l\Big)_{,k}u_l\ \ap\ \mp\mvec{k}\cdot\mvec{v}
\pm\Big[\nabla\cdot({\bf t}\cdot\mvec{v})+\nabla\cdot(\mvec{v}\mvec{p}\cdot\mvec{v})
+\p (\mvec{p}\cdot\mvec{v})\Big]\ \ap\ - \Big(t^{kl}+u^kp^l\Big)u_{l,k}.
\vspace{.3cm}\ee
The RHS of the energy balance equation \R{U1} is
\bee{aU5b}
u_lk^l\ =\ u_\alpha k^\alpha + u_4k^4\ \ap\
\mp\mvec{v}\cdot\mvec{k} \pm r, \quad r\ :=\ ck^4. 
\ee
Consequently, the energy balance equation \R{U1} becomes with \R{U5}, \R{aU5a} and
\R{aU5b}
\bee{U5b}\bullet\hspace{2cm}
\p e+\nabla\cdot\Big(\mvec{q}+e\mvec{v}\Big)+
\nabla\cdot({\bf t}\cdot\mvec{v})+\p (\mvec{p}\cdot\mvec{v})+
\nabla\cdot(\mvec{v}\mvec{p}\cdot\mvec{v})\ =\ r.\hspace{2cm}
\vspace{.3cm}\ee
Taking into account
\byy{U1d}
\nabla\cdot\Big({\bf t}\cdot\mvec{v}\Big) &=& \nabla\mvec{v}:{\bf t}^\top
+\Big(\nabla\cdot{\bf t}\Big)\cdot\mvec{v},
\\ \label{1U1d}
\p \Big(\mvec{p}\cdot\mvec{v}\Big) &=&
\p \mvec{p}\cdot\mvec{v} + \p \mvec{v}\cdot\mvec{p},
\\ \label{2U1d}
\nabla\cdot\Big(\mvec{v}\mvec{p}\cdot\mvec{v}\Big) &=& 
\Big(\nabla\cdot\mvec{v}\mvec{p}\Big)\cdot\mvec{v}
+\nabla\mvec{v}:\mvec{p}\mvec{v}, 
\eey
and taking in mind that $e$ is independent of $u^k$ according to its definition \R{J1},
and consequently also independent of $\mvec{v}$, the energy balance equation \R{U5b} results in the local rest frame 
\bee{3U1d}\bullet\hspace{2cm}
\mvec{v}\st{!}{=}\mvec{0}:\qquad
\p e+\nabla\cdot\Big(\mvec{q}+e\mvec{v}\Big)
+\nabla\mvec{v}:{\bf t}^\top
+\p \mvec{v}\cdot\mvec{p}\ =\ r.\hspace{3cm}
\ee
Consequently, $e$ is the internal energy, a fact which can be read off also from the
Gibbs equation \R{Y3}. Beyond that, \R{3U1d}  is in accordance with the balance of the
internal energy which is not based on relativistic presuppositions
\C{MU01}\footnote{$\p \mvec{v}\cdot\mvec{p}$ corresponds to
$\epsilon:\mvec{T}$ in \C{MU01}: both
terms are generated by the non-symmetry of the energy-momentum tensor}.

\subsection{The entropy balance\label{EB}}
\subsubsection{The $^A$-component of the mixture}

Starting with \R{T2} and \R{T5}, we obtain
\bee{R1}
S^{Ak}{_{,k}}\ =\ (s^Au^{Ak}){_{,k}}+s^{Ak}{_{,k}}\ =\
(s^Au^{A\alpha}){_{,\alpha}}+(s^Au^{A4}){_{,4}}
+s^{A\alpha}{_{,\alpha}}+s^{A4}{_{,4}}
=\ \sigma^A+\varphi^A.
\ee
Analogously to \R{U4}
\bee{R2}
s^{Ak}u^A_k\ =\ 0\ \longrightarrow\ s^{A4}\ \ap\ 0
\ee
is valid. In Galilean co-ordinates follows for the non-relativistic limit the usual balance
equation of the entropy
\bee{R3}\bullet\hspace{4cm}
S^{Ak}{_{,k}}\ \ap\ \nabla\cdot(\mvec{v}^As^A+\mvec{s}^A)+\p s^A\ =\ \sigma^A+\varphi^A.\hspace{4cm}
\vspace{.3cm}\ee
Especially, we consider the case of
multi-heat relaxation equilibrium and signature $(-2)$ in the sequel.
From \R{E6} and \R{K6b}$_1$
follows
\bee{R4}
\nu^A\ =\ f^A\ =\ \frac{1}{c^2}(u^A_\alpha u^\alpha + u^A_4u^4)\ \ap\ 1.
\ee
The non-relativistic limits of entropy density $s^A$, entropy flux density $\mvec{s}^A$,
entropy production $\sigma^A$ and entropy supply $\varphi^A$ follow for a component in the
mixture from \R{Z14p} to \R{T9zp} by \R{A10}.
Taking \R{U4z} into account, the entropy density of the $^A$-component \R{Z14p}
results in
\bey\nonumber
s^A &=& \frac{1}{\Theta}\Big(\mu^A\rh^A + f^A(e^A+p^A) +u_\alpha p^{A\alpha}
+u_4p^{A4}\Big)\ \ap\ 
\\ \nonumber
&\ap&\frac{1}{\Theta}\Big(\mu^A\rh^A + e^A+p^A 
-\mvec{v}\cdot\mvec{p}^A + \mvec{p}^A\cdot\mvec{v}^A\Big)\ =
\\ \label{R5}{\bullet}\hspace{4cm}
&=&\frac{1}{\Theta}\Big(\mu^A\rh^A + e^A+p^A 
+ \mvec{p}^A\cdot(\mvec{v}^A-\mvec{v})\Big).\hspace{4cm}\mbox{}
\vspace{.3cm}\eey
The diffusion flux \R{L2} becomes by taking \R{R4} into account
\bee{R6}
J^{Am}h^{A\alpha}_m\ \ap\ \rh^A(\mvec{v}^A-\mvec{v})\
=\ \mvec{J^A},\quad J^{Am}h^{A4}_m\ \ap\ 0,
\ee
and with \R{U7a}, we obtain
\byy{R7}
u_mt^{A\beta m} &=& u_\alpha t^{A\beta\alpha} + u_4t^{A\beta4}\ \ap\ -{\bf t}\cdot\mvec{v}+{\bf t}\cdot\mvec{v}\ =\ 0,
\\ \label{R8}
u_mt^{A4 m} &=& u_\alpha t^{A4\alpha} + u_4t^{A44}\ \ap\ 0.
\eey
Inserting \R{R6} to \R{R8} into \R{Z15p}, the non-relativistic limit of the entropy flux
density becomes taking \R{U4}$_3$ into account
\bee{R9}{\bullet}\hspace{6cm}
\mvec{s}^A\ =\ \frac{1}{\Theta}\Big(\mvec{q}^A-\mu^A\mvec{J}^A\Big).
\hspace{6cm}\mbox{}
\vspace{.3cm}\ee
According to \R{aU5b}, the energy supply of the $^A$-component is
\bee{R9a}
r^A\ =\ u^A_lk^{Al}\ =\ u^A_\alpha k^{A\alpha} + u^A_4k^{A4}\ \ap\
-\mvec{v}^A\cdot\mvec{k}^A + ck^{A4}
\ee
Taking \R{R4} and \R{R9a} into account, two terms of the entropy supply
\R{Z16p} become 
\bey\nonumber
\nu^A u_l^Ak^{Al} + u_ph^{Ap}_mk^{Am} &=&
\Big[\nu^Au^A_m + u_p\Big(\delta^p_m-\frac{1}{c^2}u^{Ap}u^A_m\Big)\Big]k^{Am}\ =\
\\ \nonumber
&=&\Big[(\nu^A-f^A)u^A_m + u_m\Big]k^{Am}\ =\ u_mk^{Am}\ =
\\  \nonumber
&=&\ u_\alpha k^{A\alpha}+\ u_4k^{A4}\ \ap\ 
-\mvec{v}\cdot\mvec{k}^A +ck^{A4}\ =\ 
\\ \label{R10}
&=&  -\mvec{v}\cdot\mvec{k}^A +r^A +\mvec{v}^A\cdot\mvec{k}^A.
\eey
Taking \R{R10} into account, the non-relativistic limit of the entropy supply \R{Z16p} is
in multi-heat relaxation equilibrium
\bee{R11}{\bullet}\hspace{3.2cm}
\varphi^A\ \ap\ \frac{1}{\Theta}\Big(\mu^A{^{(ex)}}\Gamma^A+
r^A +(\mvec{v}^A-\mvec{v})\cdot\mvec{k}^A\Big).\hspace{3.2cm}
\vspace{.3cm}\ee
We now will find out the different non-relativistic terms of the entropy production
\R{T9zp}. We start with the first term taking \R{L2} into account
\bee{R12}
-(J^{Am}h^{Ak}_m)_{,k}\ =-\Big[\rh^Af^A(f^Au^{Ak}-u^k)\Big]_{,k}\ \ap\
-\nabla\cdot\Big[\rh^A(\mvec{v}^A-\mvec{v})\Big]\ =\ -\nabla\cdot\mvec{J}^A.
\ee
The second term of \R{T9zp} yields by taking \R{R6} into account
\bee{R13}
-\Big(\frac{\mu^A}{\Theta^A}\Big)_{,k}J^{Am}h^{Ak}_m\ =\
-\Big(\frac{\mu^A}{\Theta^A}\Big)_{,\alpha}J^{Am}h^{A\alpha}_m
-\Big(\frac{\mu^A}{\Theta^A}\Big)_{,4}J^{Am}h^{A4}_m\ \ap
-\mvec{J}^A\cdot\nabla\Big(\frac{\mu^A}{\Theta}\Big).
\ee
Taking \R{U4}$_3$ into account, the third term of \R{T9zp} results in 
\bee{R14}
\Big(\frac{\nu^A}{\Theta^A}\Big)_{,k}q^{Ak}\ \ap\ 
\nabla\Big(\frac{1}{\Theta}\Big)\cdot\mvec{q}^A.
\ee
A term which is similar to the fourth term of \R{T9zp} is calculated in \R{aU5a}$_2$.
Consequently, we obtain
\bee{R15}
\frac{\nu^A}{\Theta^A}u^A{_{l,k}}\Big(\pi^{Akl}+u^{Ak}p^{Al}\Big)\ \ap\
-\frac{1}{\Theta}\Big(\nabla\mvec{v}^A:\mvec{\pi}^{A\top}
+(d_t\mvec{v}^A)\cdot\mvec{p}^A\Big).
\ee
Taking
\bee{R16}
\Big(\frac{1}{\Theta^A}u^mh^A_{ml}\Big)_{,k}\ =\
\Big(\frac{1}{\Theta^A}(u_l-f^Au^A_l)\Big)_{,k}
\ee
into account, the non-relativistic limit of the last term of \R{T9zp} is
\bee{R17}
\Big[\frac{1}{\Theta^A}(u_l-f^Au^A_l)\Big]_{,k}
\Big(\frac{1}{c^2}e^Au^{Ak}u^{Al}
+\frac{1}{c^2}q^{Ak}u^{Al}+t^{Akl}\Big)\ \ap\
-\nabla\Big[\frac{1}{\Theta}(\mvec{v}-\mvec{v}^A)\Big]:{\bf t}^{A\top}.
\vspace{.3cm}\ee
Inserting \R{R12} to \R{R15} and \R{R17} into \R{T9zp} results in the non-relativistic
limit of the entropy production
\bey\nonumber
\sigma^A&\ap&\frac{\mu^A}{\Theta}\Big({^{(in)}}\Gamma^A
-\nabla\cdot\mvec{J}^A\Big)-\mvec{J}^A\cdot\nabla\Big(\frac{\mu^A}{\Theta}\Big)+
\nabla\Big(\frac{1}{\Theta}\Big)\cdot\mvec{q}^A-
\\ \label{R18}
&&-\frac{1}{\Theta}\Big(\nabla\mvec{v}^A:\mvec{\pi}^{A\top}
+(d_t\mvec{v}^A)\cdot\mvec{p}^A\Big)
+\nabla\Big[\frac{1}{\Theta}(\mvec{v}^A-\mvec{v})\Big]:{\bf t}^{A\top}.
\eey
and rearranging \R{R18}, we obtain
\bey\nonumber{\bullet}\hspace{1.7cm}
\sigma^A&\ap&\frac{\mu^A}{\Theta}{^{(in)}}\Gamma^A-
\nabla\cdot\Big(\frac{\mu^A}{\Theta}\mvec{J}^A\Big)+
\nabla\Big(\frac{1}{\Theta}\Big)\cdot\mvec{q}^A-
\\ \label{R20}
&&-\frac{1}{\Theta}\Big(\nabla\mvec{v}^A:
(\mvec{\pi}^{A\top}+\mvec{p}^A\mvec{v}^A)+\p\mvec{v}^A\cdot\mvec{p}^A\Big)+
\nabla\Big[\frac{1}{\Theta}(\mvec{v}^A-\mvec{v})\Big]:{\bf t}^{A\top}.
\hspace{1.1cm}
\vspace{.3cm}\eey
The first three terms of the RHS of \R{R20} belong to the usual effects: chemical reactions,
diffusion and heat conduction. The fourth term belonging to viscosity is coupled to the
momentum balance by a modified viscosity tensor and an additional expression which
vanishes with the symmetry of the energy-momentum tensor. The last term of \R{R20}
is typical for a component of a multi-component system in non-equilibrium.

\subsubsection{The mixture}

Relativistic thermodynamics of a mixture is characterized by two items: the additivity
of the energy-momentum tensors  and of the 4-entropies of the components of the mixture resulting in the corresponding quantities of the mixture itself, \R{K11c} and \R{Q1}. 
These two settings are the starting-points for determining the (3+1)-components of the
energy-momentum tensor of the mixture in sect.\ref{COMI} and of the entropy related
quantities in sect.\ref{MHC}. Here, we are looking for their corresponding non-relativistic
limits.
\vspace{.3cm}\newline
First of all, we need the non-relativistic limits of $h^m_k$, \R{K15}, and $g^{Am}$,
\R{K15a1}:
\byy{R21}
h^\alpha_\beta &\bg& \delta^\alpha_\beta 
+ \frac{1}{c^2}\frac{v^\alpha}{\sqrt{1-(v/c)^2}}
\frac{v_\beta}{\sqrt{1-(v/c)^2}}\ \ap\ \delta^\alpha_\beta,
\\ \label{R22}
h^\alpha_4 &\bg& 0
- \frac{1}{c^2}\frac{v^\alpha}{\sqrt{1-(v/c)^2}}
\frac{c}{\sqrt{1-(v/c)^2}}\ \ap\ 0,
\\ \label{R23}
h^4_\beta &\bg& 0
+ \frac{1}{c^2}\frac{c}{\sqrt{1-(v/c)^2}}
\frac{v_\beta}{\sqrt{1-(v/c)^2}}\ \ap\ 0,
\\ \label{R24}
h^4_4 &\bg& 1 
- \frac{1}{c^2}\frac{c}{\sqrt{1-(v/c)^2}}
\frac{c}{\sqrt{1-(v/c)^2}}\ \ap\ 0,
\\ \label{R25}
g^{A\alpha}&=&
\frac{1}{c^2}\Big(u^{A\beta}h^\alpha_\beta+u^{A4}h^\alpha_4\Big)\ \ap\ 0,
\\ \label{R26}
g^{A4}&=&
\frac{1}{c^2}\Big(u^{A\beta}h^4_\beta+u^{A4}h^4_4\Big)\ \ap\ 0.
\eey
Consequently, the (3+1)-components of the mixture's energy-momentum tensor become
as follows: The third term of \R{K15c} is by taking \R{U4z}$_3$ and \R{R4} into
account
\bey\nonumber
f^A\Big(p^{A\alpha}u_\alpha + p^{A4}u_4\Big) &\bg&
f^A\Big(-\frac{\mvec{p}^A\cdot\mvec{v}}{\sqrt{1-(v/c)^2}} 
+ \frac{\mvec{p}^A\cdot\mvec{v}^A}{c}\frac{c}{\sqrt{1-(v/c)^2}}\Big)\ \ap\
\\ \label{R28}
&\ap& \mvec{p}^A\cdot(\mvec{v}^A-\mvec{v}).
\eey
Consequently, the internal energy of the mixture \R{K15c} results in
\bee{R29}\bullet\hspace{4.5cm}
{\sf e}\ \ap\ \sum_A\Big(e^A+ \mvec{p}^A\cdot(\mvec{v}^A-\mvec{v})\Big).
\hspace{4.5cm}
\vspace{.3cm}\ee
The spatial component of the energy flux density \R{K15e} is
\bey\nonumber
{\sf q}^\alpha &=& \sum_A\Big(c^2e^Af^Ag^{A\alpha} + q^{Ak}f^Ah^\alpha_k +
c^2p^{Al}g^{A\alpha}u_l + t^{Akl}h^\alpha_ku_l\Big)\ =\
\\ \label{R30}
&=& \sum_A\Big(c^2g^{A\alpha}(e^Af^A+p^{Al}u_l)+h^\alpha_k(q^{Ak}f^A+
 t^{Akl}u_l)\Big).
\eey
 Taking \R{K15a1}$_2$ and \R{K6b}$_1$ into account, special terms of \R{R30} are:
\byy{R31}
c^2g^{A\alpha} &\bg& \frac{v^{A\alpha}}{\sqrt{1-(v^A/c)^2}}-
f^A\frac{v^\alpha}{\sqrt{1-(v/c)^2}}\ \ap\ \mvec{v}^A-\mvec{v},
\\ \label{R31a}
c^2g^{A4} &\bg& \frac{c}{\sqrt{1-(v^A/c)^2}}-
f^A\frac{c}{\sqrt{1-(v/c)^2}}\ \ap\ 0,
\eey
and by taking \R{R28} into account
\bee{R32}
c^2g^{A\alpha}\Big(e^Af^A+p^{Al}u_l\Big)\ \ap\ \Big(\mvec{v}^A-\mvec{v}\Big)
\Big(e^A+\mvec{p}^A\cdot(\mvec{v}^A-\mvec{v})\Big).
\ee
By taking \R{R21}, \R{R22} and \R{U6} to \R{U7a} into account, parts of the second term of \R{R30} result in
\byy{R33}
h^\alpha_kq^{Ak}f^A &=& f^A\Big(h^\alpha_\beta q^{A\beta}
+h^\alpha_4q^{A4}\Big)\ \ap\ \mvec{q}^A,
\\ \nonumber
h^\alpha_k t^{Akl}u_l &=& h^\alpha_\beta t^{A\beta\eta}u_\eta
+ h^\alpha_\beta t^{A\beta4}u_4 +
h^\alpha_4 t^{A4\eta}u_\eta +h^\alpha_4 t^{A44}u_4\ \ap
\\ \label{R34}
&&\ap\ -{\bf t}^A\cdot\mvec{v}+{\bf t}^A\cdot\mvec{v}-{\bf 0}+{\bf 0}\ =\ {\bf 0}.
\eey
Taking \R{R32} to \R{R34} into account, the non-relativistic limit of the spatial part of
the energy flux density is
\bee{R35}\bullet\hspace{3cm}
{\sf q}^\alpha\ \ap\ \sum_A\Big[
\mvec{q}^A+\Big(e^A+\mvec{p}^A\cdot(\mvec{v}^A-\mvec{v})\Big)\Big(\mvec{v}^A-\mvec{v}\Big)\Big].\hspace{3cm}
\vspace{.3cm}\ee
The temporal component of the energy flux density \R{K15e} is according to \R{R30}
\bee{R36}
{\sf q}^4\ =\ \sum_A\Big(c^2g^{A4}(e^Af^A+p^{Al}u_l)+h^4_k(q^{Ak}f^A+
 t^{Akl}u_l)\Big).
\ee
According to \R{R31} to \R{R34},  we obtain
\bee{R37}\bullet\hspace{1.2cm}
c^2g^{A4}\Big(e^Af^A+p^{Al}u_l\Big)\ \ap\ 0,\quad
h^4_k\Big(q^{Ak}f^A+ t^{Akl}u_l\Big)\ \ap\ 0\ \longrightarrow\
{\sf q}^4\ \ap\ 0,\hspace{1.2cm}
\ee
The spatial component of the momentum flux density \R{K15g} is
\bey\nonumber
{\sf p}^\alpha &=& \sum_A\Big(e^Af^Ag^{A\alpha} +\frac{1}{c^2} q^{Ak}u_kg^{A\alpha} + p^{Al}f^Ah^\alpha_l +\frac{1}{c^2}  t^{Akl}h^\alpha_lu_k\Big)\ =
\\ \label{R39}
&=&\sum_A\Big(g^{A\alpha}(e^Af^A+\frac{1}{c^2} q^{Ak}u_k)
+h^\alpha_l ( p^{Al}f^A+\frac{1}{c^2}  t^{Akl}u_k)\Big).
\eey
According to \R{R25}, we obtain
\bee{R40}
g^{A\alpha}(e^Af^A+\frac{1}{c^2} q^{Ak}u_k)\ \ap\ {\bf 0}.
\ee
Taking \R{R21}, \R{R22} and \R{R34} into account, the second term of \R{R39}
becomes
\byy{R41}
h^\alpha_l  p^{Al}f^A\ =\ h^\alpha_\beta  p^{A\beta}f^A 
+ h^\alpha_4  p^{A4}f^A\ \ap\ \mvec{p}^A,
\\ \label{R42}
h^\alpha_l \frac{1}{c^2}  t^{Akl}u_k\ \ap\ \mvec{0},
\eey
resulting in
\bee{R43}\bullet\hspace{5.9cm}
{\sf p}^\alpha\ \ap\ \sum_A\mvec{p}^A.\hspace{5.9cm}
\vspace{.3cm}\ee
The temporal component of the momentum flux density \R{K15g} is according to
\R{R39}
\bee{R44}\bullet\hspace{1.5cm}
{\sf p}^4\ =\ \sum_A\Big(g^{A4}(e^Af^A+\frac{1}{c^2} q^{Ak}u_k)
+h^4_l ( p^{Al}f^A+\frac{1}{c^2}  t^{Akl}u_k)\Big)\ \ap\ 0,\hspace{1.5cm}
\ee
according to \R{R22}, \R{R23} and \R{R26}.
\vspace{.3cm}\newline
The spatial components of the energy momentum tensor \R{K15j} of the mixture are
\bee{R46}
{\sf t}^{\alpha\beta}\ =\ \sum_A\Big(c^2g^{A\alpha}(e^Ag^{A\beta} + p^{Al}
h^\beta_l)+h^\alpha_k(q^{Ak}g^{A\beta} + t^{Akl}h^\beta_l)\Big).
\ee
The first term is by taking \R{R31}, \R{R25}, \R{R21} and  \R{R22} into account
\bee{R47}
c^2g^{A\alpha}(e^Ag^{A\beta} + p^{Al}h^\beta_l)\ \ap\
(\mvec{v}^A-\mvec{v})\mvec{p}^A.
\ee
Taking \R{R21} and \R{R22} into account, the second term of \R{R46} becomes
\bey\nonumber
h^\alpha_k(q^{Ak}g^{A\beta} + t^{Akl}h^\beta_l) &=&
h^\alpha_\eta q^{A\eta}g^{A\beta} + h^\alpha_4q^{A4}g^{A\beta}+
\\ \nonumber
&+&h^\alpha_\eta t^{A\eta \xi}h^\beta_\xi + h^\alpha_4 t^{A4\xi}h^\beta_\xi
+ h^\alpha_\eta t^{A\eta 4}h^\beta_4 + h^\alpha_4 t^{A44}h^\beta_4\ \ap\
\\ \label{R48}
&\ap& {\bf t}^A.
\eey
Consequently, the non-relativistic limit of the stress tensor \R{R46} of the mixture is
\bee{R49}\bullet\hspace{4.5cm}
{\sf t}^{\alpha\beta}\ \ap\ \sum_A\Big({\bf t}^A +
(\mvec{v}^A-\mvec{v})\mvec{p}^A\Big).\hspace{4.5cm}
\vspace{.3cm}\ee
The three other components of the energy momentum tensor of the mixture are as
follows:
\byy{R50}
{\sf t}^{\alpha 4} &=& \sum_A\Big(c^2g^{A\alpha}(e^Ag^{A4} + p^{Al}
h^4_l)+h^\alpha_k(q^{Ak}g^{A4} + t^{Akl}h^4_l)\Big),
\\ \label{R51}
{\sf t}^{4\beta} &=& \sum_A\Big(c^2g^{A4}(e^Ag^{A\beta} + p^{Al}
h^\beta_l)+h^4_k(q^{Ak}g^{A\beta} + t^{Akl}h^\beta_l)\Big),
\\ \label{R52}
{\sf t}^{44} &=& \sum_A\Big(c^2g^{A4}(e^Ag^{A4} + p^{Al}
h^4_l)+h^4_k(q^{Ak}g^{A4} + t^{Akl}h^4_l)\Big).
\eey
Taking \R{R31a}, \R{R23} and \R{R24} into account, \R{R50} to \R{R52} result in
\byy{R53}
{\sf t}^{\alpha 4} &\ap& \sum_A\Big(h^\alpha_k(q^{Ak}g^{A4} + t^{Akl}h^4_l)\Big),
\\ \label{R54}
{\sf t}^{4\beta} &\ap& \sum_A\Big(h^4_k(q^{Ak}g^{A\beta} + t^{Akl}h^\beta_l)\Big),
\\ \label{R55}
{\sf t}^{44} &\ap& \sum_A\Big(h^4_k(q^{Ak}g^{A4} + t^{Akl}h^4_l)\Big).
\eey
Taking \R{R31a} and \R{R23} into account, we obtain
\bee{R56}\bullet\hspace{4.3cm}
{\sf t}^{\alpha 4}\ \ap\ 0,\quad
{\sf t}^{4\beta}\ \ap\  0,\quad
{\sf t}^{44}\ \ap\ 0.\hspace{4.3cm}
\vspace{.3cm}\ee
The non-relativistic limit of the pressure of the mixture \R{K15j1} is
\bee{R57}\bullet\hspace{5.2cm}
{\sf p}\ \ap\ -\frac{1}{3}\sum_A t^{Ak}_k\ =\ \sum_A p^A.\hspace{5.2cm}
\ee

\section{Summary}

A multi-component system is formed by its components which are characterized by own individual
quantities, such as velocity, density, chemical potential, stress tensor, temperature, heat flux and
entropy flux densities, entropy production and supply and further items. All these individual
quantities determine those of the multi-component system which is described as a mixture.  
Individual temperatures of the components result in multi-heat relaxation towards the corresponding
equilibrium generating a common temperature of all components and the mixture. A temperature
of the mixture in multi-heat relaxation non-equilibrium depends on the used thermometer and
cannot be defined unequivocally.
\vspace{.3cm}\newline
Starting out with the rest mass densities of the components of the multi-component system,
the mass flux densities of the components are defined by introducing their different 4-velocities.
The mixture of the components is characterized by several settings. The first one is the additivity
of the component's mass flux densities to the mass flux density of the mixture. In combination with
the mixture axiom, this setting allows to define mass density and 4-velocity of the mixture and the
diffusion fluxes of the components. The non-symmetric energy-momentum tensor of one component interacting
with the mixture is introduced, and its (3+1)-split together with the component's mass and diffusion
flux densities are generating the entropy identity \C{MUBO}. The exploitation of the entropy identity requires additional settings: the entropy density, flux and supply. These settings result in physical
interpretations of entropy density, flux and supply. The entropy production follows from the entropy
identity which restricts possible arbitrariness of defining. 
\vspace{.3cm}\newline
The use of the entropy identity introduces so-called accessory variables which are Lagrange
parameters concerning the constraints taken into considerstion. These are temperature,
chemical potential and an additional non-equilibrium variable which characterizes the considered
component to be a part of the mixture. Beside the classical irreversible processes --diffusion,
chemical reactions, heat conduction and friction-- an additional irreversible process --multi-heat
relaxation-- appears due to
the embedding of the considered component into the mixture. Different from the classical case, the
mass production term, the heat flux density and the viscous tensor are modified by so-called effective
quantities.
\vspace{.3cm}\newline
Equilibrium is defined by equilibrium conditions which are divided into
necessary and supplementary ones \C{MUBO,MUBO1,BOCHMU}. The necessary equilibrium conditions are
given by vanishing entropy production, vanishing entropy flux density and vanishing
entropy supply. Supplementary equilibrium conditions are: vanishing diffusion flux densities,
vanishing component time derivatives\footnote{except that of the 4-velocity} and vanishing of the
mass production terms. Presupposing these equilibrium conditions, we obtain: all components have
the same 4-velocity, all heat flux densities are zero, the power as well as the divergence of the
4-velocity of each component vanish, and the viscous tensor is perpendicular to the velocity gradient.
\vspace{.3cm}\newline
The corresponding free component is defined by undistinguishable component
indices\footnote{that is not the mixture which is a multi-component system}. This 1-component
system represents the easiest classical case serving as a test, if the interacting component in the
mixture is correctly described. The vanishing of the entropy production in equilibrium is shortly
investigated: the so-called Killing relation of the vector of 4-temperature is neither a necessary nor
a sufficient condition for equilibrium. Also the statement that materials are perfect in equilibrium
cannot be confirmed.
\vspace{.7cm}\newline
{\bf Ackowledgement} Discussions with  H.-H. v. Borzeszkowski are gratefully acknowledged.

\section{Appendices}
\subsection{Rest mass densities\label{RMD}}

Consider two frames, ${\cal B}^A$ and ${\cal B}^B$. ${\cal B}^A$ is the rest frame of the
$^A$-component and ${\cal B}^B$ that of the $^B$-component. The corresponding rest mass
densities are
\bee{A1}
\mbox{rest densities:}\qquad\rh^A\ \mbox{in}\ {\cal B}^A,\qquad
\rh^B\ \mbox{in}\ {\cal B}^B.
\ee
By definition, the rest mass densities do not depend on the frame, that means, the rest mass
densities are {\em relativistic invariants} which should not be confused with the measured densities
in non-resting frames
\bee{A3}
{\cal B}^B:\quad\rh^A_B\ =\ \frac{\rh^A}{1-v^2_{AB}/c^2},\qquad\quad
{\cal B}^A:\quad\rh^B_A\ =\ \frac{\rh^B}{1-v^2_{BA}/c^2},\qquad v_{AB}\ =\ -v_{BA}.
\ee
Here $\rh^A_B$ is the density of the $^A$-component in the rest frame of the $^B$-component,
and $v_{AB}$ is the translational 3-velocity of ${\cal B}^A$ in the frame ${\cal B}^B$. These
densities are out of scope in this paper. 
\vspace{.3cm}\newline
We now consider \R{K6b}$_1$ in the rest frame ${\cal B}_O$ of the mixture which is defined by
$u^k_O = (0,0,0,c)$. Consequently, we obtain
\bee{A4a}
f^A_O\ =\ \frac{1}{c^2}u^A_{4O}c\ =\ 
\frac{1}{c}\frac{c}{\sqrt{1-v^2_{AO}/c^2}}.
\ee
Inserting \R{A4a} into \R{K6b}$_2$ results in the mass density of the mixture in its rest frame 
\bee{A4b}
{\cal B}_O:\quad
\rh_O\ =\ \sum_A f^A_O\rh^A_O\ =\ 
\sum_A\frac{1}{\sqrt{1-v^2_{AO}/{c^2}}}\frac{\rh^A}{1-v^2_{AO}/c^2}\ =\ 
\sum_A\frac{\rh^A}{(1-v^2_{AO}/c^2)^{3/2}}.
\ee
The same result is obtained, if \R{K5}$_3$ is written down for the rest system of the mixture.

\subsection{Example: Uniform component velocities\label{UCV}}

If there exists a common rest frame ${\cal B}^0$ for all
$^A$-components 
\bee{aK11}
u^A_k\ \doteq\ u^0_k,\quad \wedge A.
\ee
According to \R{K5}$_3$, we obtain
\bee{fK11}
\rh u_k\ =\ u^0_k\sum_ A\rh^A\quad\longrightarrow\quad
\rh c^2\ =\ u^ku^0_k\sum_ A\rh^A\ \wedge\ \rh u_ku^{0k}\ =\ c^2\sum_ A\rh^A,
\ee
and with \R{K6b}$_1$ follows
\bee{gK11}
\rh c^2\ =\ c^2f^0\sum_ A\rh^A\ \wedge\ \rh c^2f^0\ =\ c^2\sum_ A\rh^A
\quad\longrightarrow\quad(f^0)^2\ =\ 1,
\ee
resulting in
\bee{cK11}
f^0\ =\ \pm 1.
\ee
We obtain from \R{K6b}$_2$
\bee{dK11}
\varrho\ =\ f^0\sum_a\varrho^A\ =\ \pm \sum_a\varrho^A
\quad\longrightarrow\quad
f^0\ =\ +1,
\ee
and taking \R{fK11}$_1$ into account
\bee{eK11}
u_k\ =\ u^0_k.
\ee
As expected, the 4-velocity of the mixture is identical with the
uniform component velo\-ci\-ties.

\subsection{Stoichiometric equations\label{SE}}

The system of the relativistic stoichiometric equations runs as follows
\byy{A5}
\sum_A\nu_\alpha^A M^A_0\ =\ 0,\hspace{4.5cm}
\\ \nonumber
\mbox{component index: }A=1,2,...,Z,\qquad
\mbox{reaction index: }\alpha=1,2,...,\Omega. 
\eey
The stoichiometric coefficients $\nu_\alpha^A$ are scalars, and the partial rest mole mass $M^A_0$
is defined using the scalar mole number $n^A$ and the mole concentration $\zeta^A$ of the
$^A$-component  
\bee{A6}
M^A_0\ :=\ \frac{m^A_0}{n^A}\ =\ \frac{V_0}{n^A}\rh^A\ =\ \frac{\rh^A}{\zeta^A},\quad
\zeta^A\ :=\ \frac{n^A}{V_0}
\ee
according to \R{A1} and \R{A2}. The stoichiometric coefficients $\nu_\alpha^A$ are determined
by the partial rest mole masses $M^A_0,\ A=1,2,...,Z,$ before and after the $\alpha th$ reaction.
\vspace{.3cm}\newline
The time derivative of the mole number is determined by the reaction velocities
$\st{\td}{\xi}_\alpha$
\bee{A7}
\st{\td}{n}{^A}\  =\ \sum_\alpha\nu_\alpha^A\st{\td}{\xi}_\alpha.
\ee
Multiplication with $M^A_0$ results by use of \R{A5} in
\bee{A8}
M^A_0\st{\td}{n}{^A}\ =\ \sum_\alpha\nu_\alpha^A M^A_0\st{\td}{\xi}_\alpha
\ \longrightarrow\ \sum_A M^A_0\st{\td}{n}{^A}\ =\ 0.
\vspace{.3cm}\ee
Starting out with the physical dimensions
\bee{A9}
[\nu_\alpha^A]\ =\ mol,\ \ [n]\ =\ mol,\ \ [M^A_0]\ =\ \frac{kg}{mol},\ \ 
[\zeta^A]\ =\ \frac{mol}{m^3},\ \ [\st{\td}{\xi}_\alpha]\ =\ \frac{1}{s},
\ee
that of the mass production term in the first row of \R{T9z} is evidently
\bee{Z13a}
[{^{(in)}}\Gamma^A]\ =\ \frac{kg}{m^3s}.
\ee
A comparison with
\bee{Z13b}
[M^A_0\st{\td}{n}{^A}]\ =\ \frac{kg}{s}
\ee
shows that ${^{(in)}}\Gamma^A$ is the density which belongs to the mass production \R{A8}.
Because
according to \R{A2}, all rest mass densities are referred to the relativistic invariant $V_0$,
we obtain from \R{Z13b} and \R{Z13a} with \R{A8}
\bee{Z13c}
{^{(in)}}\Gamma^A\ =\ \frac{1}{V_0}\sum_\alpha\nu_\alpha^A M^A_0\st{\td}{\xi}_\alpha\
\longrightarrow\ \sum_A{^{(in)}}\Gamma^A\ =\ 0.
\ee

\subsection{Remark: Constitutive equations and the 2$^{\rm nd}$ Law}

Up to here, a special material was not taken into account: all considered relations are
valid independently of the material which is described by constitutive equations
supplementing the balance equations. Especially, the entropy productions \R{T9zp} of
the $^A$-component and \R{Q10} of the mixture are not specified for particular
materials. There are different possibilities for introducing constitutive
equations\footnote{as an ansatz, or better by construction procedures \C{LIU,MUEH}}.
Because constitutive equations are not in the center of our
considerations, we restrict ourselves on the easiest ansatz which only serves for
elucidation.
\vspace{.3cm}\newline
The entropy production of the $^A$-component is a sum of two-piece products
whose factors are so-called fluxes and forces. According to \R{T9z}, the fluxes are
\bey\nonumber
{\cal Y}^A\ =\ 
\Big\{{^{(in)}}\Gamma^A-\Big(J^{Am}h^{Ak}_m\Big)_{,k},\ 
\Big(J^{Am}h^{Ak}_m\Big)_{,k},\ 
q^{Ak},\ \pi^{Akl}+u^{Ak}p^{Al}, 
\\ \label{O24}
H^{ABk},\ 
\frac{1}{c^2}e^Au^{Ak}u^{Al}+\frac{1}{c^2}q^{Ak}u^{Al}+t^{Akl}\Big\},
\eey
and the corresponding forces are
\byy{O25}
{\cal X}^A\ =\ 
\Big\{\frac{\mu^A}{\Theta^A},\ \Big(\frac{\mu^A}{\Theta^A}\Big)_{,k},\ 
\Big(\frac{f^A}{\Theta^A}\Big)_{,k},\ \frac{f^A}{\Theta^A}u^A{_{l,k}},\ 
\Big(\frac{1}{\Theta^A}-\frac{1}{\Theta^B}\Big)_{,k},\ 
\Big(\frac{1}{\Theta^A}u^mh^A_{ml}\Big)_{,k}\Big\}.
\vspace{.3cm}\eey
The entropy production density \R{T9z} writes symbolically
\bee{O26}
\sigma^A\ =\ {\cal Y}^A\cdot{\cal X}^A.
\ee
The material is described by the dependence of the fluxes on the forces, by the
constitutive
equations
\bee{O27}
{\cal Y}^A\ =\ {\bf F^A}({\cal X}^A),
\ee
resulting in
\bee{O28}
\sigma^A\ =\ {\bf F^A}({\cal X}^A)\cdot{\cal X}^A
\qquad\longrightarrow\qquad
^\dm\,\sigma\ =\ \sum_A {\bf F^A}({\cal X}^A)\cdot{\cal X}^A\ \geq\ 0.
\ee
The inequality is caused by the Second Law which states that the entropy production of
the mixture is not negative for all events after having inserted the constitutive equations
into the general expression \R{Q10}. The entropy production of an $^A$-component
\R{T9zp} is not necessarily positive semi-definite. There are different methods for
exploiting the dissipation inequality \R{O28}$_2$ \C{TRIPACIMU,MUTRIPA} which are beyond this paper.

\end{document}